\documentclass[twocolumn,11pt]{infocom}
\usepackage{cite}
\usepackage{graphicx}
\usepackage{psfrag}
\usepackage{epsfig}
\graphicspath{{.}}

\begin{document}

\title{A statistical approach to the traceroute-like exploration of 
networks: theory and simulations}

\author{Luca Dall'Asta, Ignacio Alvarez-Hamelin, Alain Barrat,
Alexei V{\'a}zquez, Alessandro Vespignani 
\thanks{L. Dall'Asta, I. Alvarez-Hamelin, A. Barrat, A. Vespignani are  
with L.P.T., B\^atiment 210, Universit\'e de Paris-Sud, 91405 ORSAY Cedex France.}
\thanks{A. V{\'a}zquez is with Department of Physics,
University of Notre Dame, Notre Dame, IN 46556, USA.}
\thanks{I. Alvarez-Hamelin is also with Facultad de Ingenier\'{\i}a, Universidad de Buenos Aires, Paseo Col\'on 850, C 1063 ACV Buenos Aires, Argentina.}
\thanks{A. Vespignani is also with School of Informatics and Department of Physics, University of Indiana, Bloomington, IN 47408, USA.}
}

\maketitle
\begin{abstract}
Mapping the Internet generally consists in  sampling the network 
from a limited set of sources by using \texttt{traceroute}-like probes. 
This methodology, akin to the merging of different spanning trees to 
a set of destinations, has been
argued to introduce uncontrolled sampling biases that might produce
statistical properties of the sampled graph which sharply differ from
the original ones\cite{crovella,clauset,delos}.
Here we explore these biases and provide a statistical analysis of
their origin. We derive a mean-field analytical approximation for the 
probability of edge and vertex detection that exploits the role of 
the number of sources and targets and allows us to relate the global
topological properties of the underlying network with the statistical
accuracy of the sampled graph. In particular we find
that the edge and vertex detection probability is depending on 
the {\em betweenness centrality} of each element.  This allows us to 
show that shortest path routed
sampling provides a better characterization of underlying
graphs with scale-free topology. We complement the analytical 
discussion with a throughout numerical investigation of simulated
mapping strategies in different network models.  
We show that sampled graphs provide a fair qualitative 
characterization of the statistical properties of the original 
networks in a fair range of different strategies and exploration 
parameters. The numerical study also allows the identification of
intervals of the exploration parameters that optimize the fraction 
of nodes and edges discovered in the sampled graph. This finding
might hint the steps toward more efficient  mapping 
strategies.
\end{abstract} 


\section{Introduction}

A significant research and technical challenge in the study 
of large information networks is related to the lack of highly 
accurate maps providing information on their basic topology.
This is mainly due to the dynamical nature 
of their structure and to the lack of any centralized control 
resulting in a self-organized growth and evolution of these systems.
A prototypical example of this situation is  faced in the case of 
the physical Internet. The topology of the Internet can be investigated
at different granularity levels such as the router and Autonomous
System (AS) level, with  the final aim of obtaining an abstract 
representation where the set of routers or ASs and their physical 
connections (peering relations) are the vertices 
and edges of a graph, respectively. In the absence of accurate 
maps, researchers rely on a general strategy that consists 
in acquiring local views of the network from several 
vantage points and merging these views in order to get a presumably
accurate global map.
Local views are obtained by evaluating a
certain number of paths to different destinations by using specific 
tools such as \texttt{traceroute} or by the analysis of BGP tables. 
At first approximation these processes amount to the collection of
shortest paths from a source node to a set of target nodes, obtaining
a partial spanning tree of the network. The merging of several of
these views provides the map of the Internet from which the statistical
properties of the network are evaluated. 

By using this strategy a number of research groups have generated 
maps of the Internet~\cite{nlanr,caida,asdata,scan,lucent}, that have 
been used for the statistical characterization of the network
properties. Defining $\mathcal{G}=(V,E)$
as the sampled graph of the Internet with $N=|V|$ vertices and $|E|$ edges, 
it is quite intuitive that the Internet is a {\em sparse} graph
in which the number of edges is much lower than in a complete graph;
i.e. $|E| \ll N(N-1)/2$. Equally important is the fact that the average
distance, measured as the shortest path, between vertices is very
small. This is the so called {\em small-world} property, that is
essential for the efficient functioning of the network.  
Most surprising is the evidence of a power-law relationship between 
the frequency of vertices and their degree $k$ defined as the number 
of edges linking each vertex to its neighbors. 
Namely, the probability that any vertex in the graph has degree $k$ is well
approximated by $P(k)\sim k^{-\gamma}$ with $2\le \gamma\le 2.5$  
\cite{faloutsos}. Evidence
for the heavy-tailed behavior of the degree distribution has been 
collected in several other studies at the router and AS level 
\cite{mercator,broido,calda,us02,chen02} and have generated a large
activity in the field of network modeling and characterization 
\cite{brite,inet,mdbook,baldi,psvbook}.

While \texttt{traceroute}-driven strategies are very flexible and 
can be feasible for extensive use, the obtained maps are undoubtedly
incomplete. Along with technical problems such as the instability of
paths  between routers and  interface resolutions \cite{burch99},
typical mapping projects are run from relatively small sets of sources
whose combined  views are missing a considerable number of edges
and vertices \cite{chen02,willinger}. In particular, the various
spanning trees are specially missing the lateral connectivity of 
targets and sample more frequently nodes and links which are closer 
to each source, introducing spurious effects that might seriously
compromise the statistical accuracy of the sampled graph.
These {\em sampling biases}
have been explored in numerical experiments of synthetic graphs 
generated by different algorithms\cite{crovella,clauset,delos}. 
Very interestingly, it has been shown that apparent degree 
distributions with heavy-tails may be 
observed even from regular topologies such as in the classic 
Erd{\"o}s-R{\'e}nyi graph model\cite{crovella,clauset}. 
These studies thus point out that the evidence obtained from the 
analysis of the Internet sampled graphs might be insufficient to draw
conclusions on the topology of the actual Internet network. 

In this work we tackle this problem by performing a mean-field
statistical analysis and extensive numerical
experiments of shortest path routed sampling in different networks models.
We find an approximate expression for the
probability of edges and vertices to be detected that exploits the
dependence upon the number of sources, targets and the topological
properties 
of the networks. 
This expression allows the understanding 
of the qualitative behavior of the efficiency of the exploration
methods by changing the number of probes imposed to the
graph. Moreover, the analytical study provides a general understanding 
of which kind of topologies yields the most accurate sampling.
In particular, we show that the map accuracy depends on
the underlying network {\em betweenness centrality} distribution; 
the broader the distribution
the higher the statistical accuracy of the sampled graph. 

We substantiate our analytical finding with a throughout analysis 
of maps obtained varying the number of source-target pairs 
on networks models with different topological properties. 
The results show that single source mapping processes 
face serious limitations in that also the targeting of the whole
network results in a very partial discovery of its
connectivity. On the contrary, the use of multiple sources 
promptly leads to a consistent increase in the accuracy 
of the obtained maps where the  statistical
degree distributions are qualitatively discriminated also at low
values of target density. A detailed discussion
of the behavior of the degree distribution and other statistical 
quantities as a function of target and sources is provided for 
sampled graphs with different topologies and compared with the insight 
obtained by analytical means. 

We also inspect quantitatively the 
portion of discovered network in different mapping process
imposing the same density of probes to the network. We find
the presence of a region of low efficiency (less nodes and edges
discovered) depending on the relative proportion of sources and
targets. Furthermore, the analysis of the optimal range of sources and
targets for the estimate of the network average degree and clustering 
indicates a different parameters region. This finding calls for a ``trade-off''
between the accuracy in the observation of different quantities and
hints to possible optimization procedures in the \texttt{traceroute}-driven
mapping of large networks.

\section{Related work}

A certain number of works have been devoted to the study of sampled
graphs obtained by shortest path probing procedures, and to the assessment
of their accuracy. We present a short survey of the works which are
related to ours.

Work by Lakhina et al. \cite{crovella} has shown that power-law like
distributions can be obtained for subgraphs of Erd\"os-R\'enyi random graphs
when the subgraph is the result of a \texttt{traceroute} exploration with
relatively few sources and destinations. They discuss the origin of
these biases and the effect of the distance between source and target in
the mapping process.

In a recent work \cite{clauset},  Clauset and Moore
have studied analytically the single source probing to
all possible destinations of an Erd\"os-Renyi random graph 
with average degree $\overline{k}$. In agreement with the 
numerical study of Lakhina et al. \cite{crovella} they have found that 
the connectivity distribution of
the obtained spanning tree displays a power-law behavior $k^{-1}$, with
an exponential cut-off setting in at a characteristic degree 
$k_c\sim\overline{k}$.

In Ref.~\cite{delos}, Petermann and De Los Rios have 
studied a \texttt{traceroute}-like procedure on
various examples of scale-free graphs, showing that, in the case of a
single source, power-law distributions with underestimated
exponents are obtained. Analytical estimates of the measured exponents as
a function of the true ones were also derived.
Finally, in a recent preprint appeared during the completion of our work,
Guillaume and Latapy \cite{latapy} report about the shortest-paths
explorations of synthetic graphs, comparing properties of the
resulting sampled graph with those of the original network. 
The exploration is made using level plots for the proportion of 
discovered nodes and edges in the graph as a function of the
number of sources and targets, giving also hints for optimal placement
of sources and targets.
All these pieces of work make clear the relevance of   
determining up to which extent the topological properties observed 
in sampled graphs are representative of that of 
the real networks.

\section{Modeling the \texttt{traceroute} discovery of unknown networks}
\begin{figure}[t]
\begin{center}
\includegraphics[width=6cm]{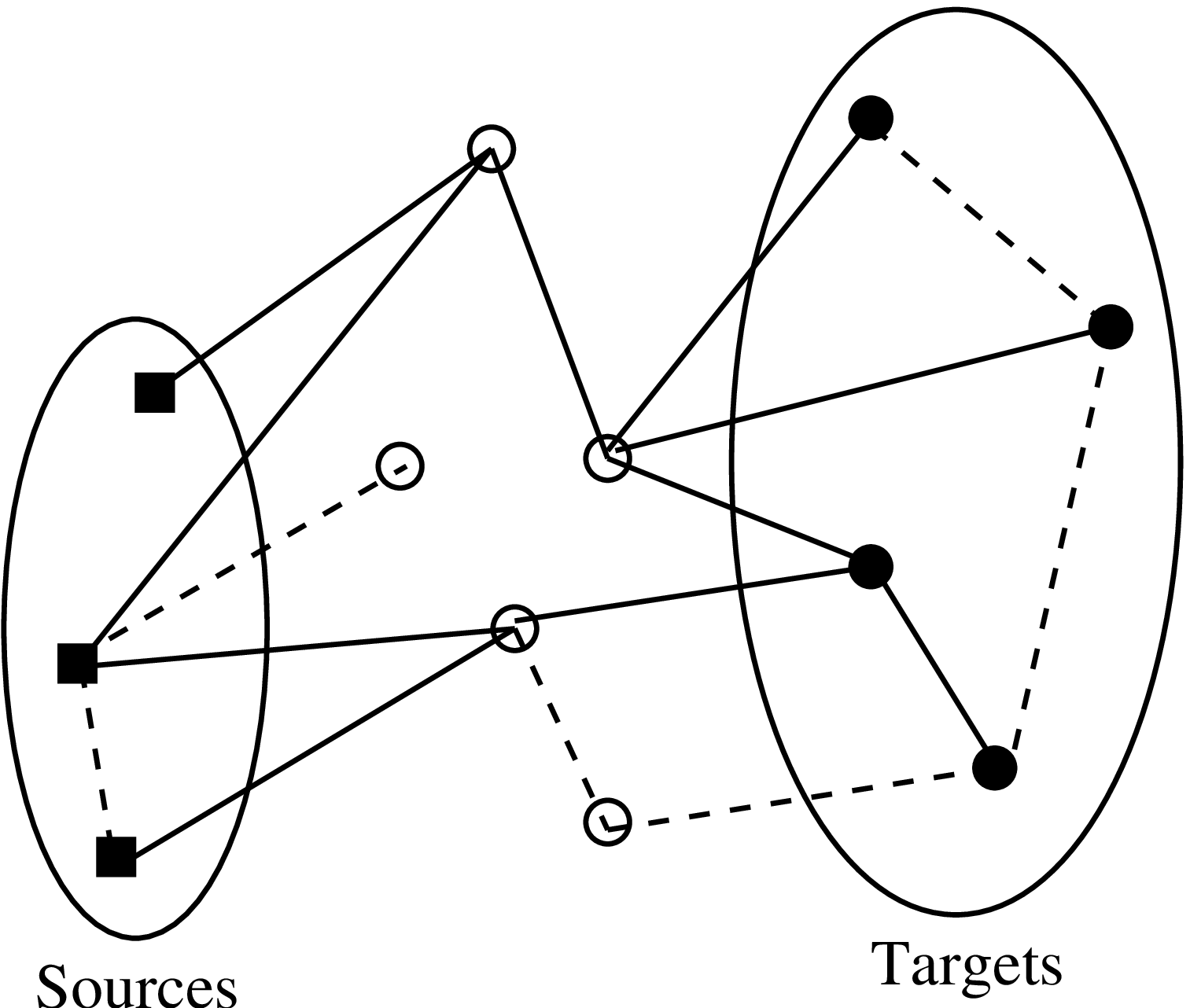}
\end{center}
\caption{Illustration of the \texttt{traceroute}-like procedure. 
Shortest paths between the set of sources and the set of destination 
targets are
discovered (shown in full lines) while other edges are not found
(dashed lines). Note that not all shortest paths are found since the
``Unique Shortest Path'' procedure is used. }
\label{fig:1}
\end{figure}

In a typical \texttt{traceroute} study, a set of active sources 
deployed in the network run \texttt{traceroute} probes to a set of 
destination nodes. Each probe collects information
on all the nodes and edges traversed along the path connecting the source to
the destination, allowing the discovery of the network \cite{burch99}. 
By merging the information collected on each path it is then possible
to reconstruct a partial map of the network (see Fig.\ref{fig:1}). 
More in detail, the edges
and the nodes discovered by each probe will depend on the metric $\mathcal{M}$
used to decide  the path between a pair of nodes. While in the
Internet many factors, including commercial agreement and
administrative routing policies, contribute to determine the
actual path, it is clear that to a first approximation the route
obtained by \texttt{traceroute}-like probes is the shortest path between
the two nodes. This 
assumption, however, is not sufficient for a proper 
definition of a \texttt{traceroute}
model in that equivalent shortest paths between two nodes may exist.
In the presence of a degeneracy of shortest paths we must therefore
specify the metric $\mathcal{M}$ by providing a resolution algorithm for
the selection of shortest paths.

For the sake of simplicity we can define three selection
mechanisms defining different $\mathcal{M}$-paths  
that may account for some of the features encountered in Internet
discovery:
\begin{itemize}
\item Unique Shortest Path (USP) probe. In this case the shortest
path route selected between a node $i$ and the destination target $T$ 
is always the same independently of the source $S$ (the path being
initially chosen at random among all the equivalent ones). 

\item Random Shortest Path (RSP) probe. The shortest path between
any source-destination pair is chosen randomly among the set of
equivalent shortest paths. This might mimic different peering
agreements that make independent the paths among couples of nodes.

\item All Shortest Paths (ASP) probe. The metric discovers all the 
equivalent shortest paths between source-destination pairs. This might
happen in the case of probing repeated in time (long time
exploration), so that back-up paths and equivalent paths are 
discovered in different runs.  
\end{itemize}
Actual \texttt{traceroute} probes contain a mixture of the three
mechanisms defined above. We do not attempt, however, to account for all
the subtleties that real studies encounters, i.e. IP routing, BGP
policies, interface resolutions and many others. 
Each \texttt{traceroute} probe provides a test of the possible biases and
we will see that the different metrics have only little influence on
the general picture emerging from our results. 
On the other hand, it is intuitive to recognize that the USP metric 
represents the worst case scenario since, among the three different methods,
it yields the minimum number of discoveries. For this
reason, if not otherwise specified, we will report the USP
data to illustrate the general features of our synthetic exploration.

More formally, the experimental setup for our simulated \texttt{traceroute}
mapping is the following. Let $G=(V,E)$ be a sparse undirected graph
with vertices (nodes) $V=\{1,2\cdots,N\}$ and edges (links) 
$E=\{i,j\}$. Then let us define the sets of vertices $\mathcal{
S}=\{i_1,i_2,\cdots,i_{N_S}\}$ and $\mathcal{
T}=\{j_1,j_2,\cdots,j_{N_T}\}$ specifying
the random placement of $N_S$ sources and $N_T$ destination targets.
For each ensemble of source-target pairs $\Omega=\{ \mathcal{S}, \mathcal{T} \}$, 
we compute with the metric $\mathcal{M}$  the path connecting 
each source-target pair. The sampled graph $\mathcal{G}=(V^*,E^*)$ 
is defined as the set of vertices $V^*$ (with $N^*=|V^*|$) and edges 
$E^*$ induced by 
considering the union of all the $\mathcal{M}$-paths connecting 
the source-target pairs. 
The sampled graph is thus analogous to the maps obtained from real 
\texttt{traceroute} sampling of the Internet.

In our study the  parameters of interest are the density $\rho_T=N_T/N$
and $\rho_S=N_S/N$ of targets and sources.
In general, \texttt{traceroute}-driven studies run from a relatively
small number of sources to a much larger set of destinations.
For this reason, in many cases it is appropriate to work with 
the density of targets $\rho_T$ while still considering 
$N_S$ instead of the corresponding density. Indeed, it is 
clear that while $100$ targets may represent a fair probing of a 
network composed by $500$ nodes, this number would be clearly 
inadequate in a network of $10^6$ nodes. On the contrary, 
the density of targets $\rho_T$ allows us to compare mapping 
processes on networks with different sizes by
defining an intrinsic percentage of targeted vertices.
In many cases, as we will see in the next sections, an appropriate 
quantity representing the level of sampling of the networks is  
\begin{equation}
\epsilon= \frac{N_S N_T} {N} = \rho_T N_S,
\end{equation}
that measures the density of probes imposed to the system. In real
situations it represents the density of \texttt{traceroute} probes in the network
and therefore a measure of the load provided to the network by the
measuring infrastructure.

In the following, our aim is to evaluate to which extent the
statistical properties of the sampled graph $\mathcal{G}$ 
depend on the parameters of our experimental setup and are 
representative of the properties of the underlying graph $G$.

\section{Mean-field theory of the discovery bias}
\label{stat}

We begin our study by presenting a mean-field statistical
analysis of the simulated \texttt{traceroute} mapping. Our aim is to
provide a statistical estimate for the probability of edge and node
detection as a function of $N_S$, $N_T$ and the topology of the
underlying graph. 

For each set $\Omega=\{ \mathcal{S}, \mathcal{T} \}$ we can define the
quantities 
\begin{equation}
\sum_{t=1}^{N_T} \delta_{i,j_t}= \left\{
\begin{array}{ll}
1 & \mbox{ if vertex $i$ is a target;}\\
0 & \mbox{ otherwise,}
\end{array}
\right.
\end{equation}
\begin{equation}
\sum_{s=1}^{N_S} \delta_{i,i_s}= \left\{
\begin{array}{ll}
1 & \mbox{ if vertex $i$ is a source;}\\
0 & \mbox{ otherwise,}
\end{array}
\right.
\end{equation}
where $\delta_{i,j}$ is the Kronecker symbol.
These quantities tell us if any given node $i$ belongs to the
set of sources or targets, and obey the sum rules 
$\sum_i\sum_{t=1}^{N_T} \delta_{i,j_t}=N_T$ and 
$\sum_i \sum_{s=1}^{N_S} \delta_{i,i_s}=N_S$. Analogously, we
define the quantity $\sigma_{i,j}^{(l,m)}$ that assumes the value $1$ 
if the edge $(i,j)$ belongs to the $\mathcal{M}$-path 
between nodes $l$ and $m$, and $0$ otherwise. 
By using the above definitions, the
indicator function that a given edge $(i,j)$ will be discovered and belongs to
the sampled graph is given by 
\begin{equation}
\pi_{i,j}=1-
\prod_{l\neq m}\left(
1 - 
\sum_{s=1}^{N_S} \delta_{l,i_s} 
\sum_{t=1}^{N_T} \delta_{m,j_t} 
\sigma_{i,j}^{(l,m)}
\right).
\end{equation}
In the case of a given set $\Omega=\{ \mathcal{S}, \mathcal{T} \}$, the 
discovery indicator function is simply $\pi_{i,j}=1$ if the edge $(i,j)$ 
belongs to at least one of the $\mathcal{M}$-paths connecting the 
source-target pairs, and $0$ otherwise.
While the above exact expression does not lead us too far in the
understanding of the discovery probabilities, it is interesting to
look at the process on a statistical ground by studying the average over 
all possible realizations of the set $\Omega=\{ \mathcal{S}, \mathcal{T} \}$.
By definition we have that 
\begin{equation}
\left\langle \sum_{t=1}^{N_T} \delta_{i,j_t} \right\rangle
= \rho_T~~~\mbox{and}~~
\left\langle \sum_{s=1}^{N_S} \delta_{i,i_s} \right\rangle
= \rho_S,
\label{aver}
\end{equation}
where $\left\langle\cdots\right\rangle$ identifies the average over
all possible deployment of sources and targets $\Omega$. 
These equalities simply
state that each node $i$ has, on average, a probability to be a source
or a target that is proportional to their respective densities. 
In the following, we will make use of an uncorrelation assumption
that allows an explicit approximation
for the discovery probability. The assumption consists in 
neglecting correlations originated by the position of sources and
targets on the discovery probability by different paths. 
While this assumption does not provide an exact treatment for the
problem it generally conveys a qualitative understanding of the 
statistical properties of the system. 
In this approximation, the average discovery probability of an edge is 
\begin{eqnarray}
\nonumber&&\left\langle\pi_{i,j}\right\rangle=1-
\left\langle\prod_{l\neq m}\left(
1 - 
\sum_{s=1}^{N_S} \delta_{l,i_s} 
\sum_{t=1}^{N_T} \delta_{m,j_t} 
\sigma_{i,j}^{(l,m)}
\right)\right\rangle\\
&&~~~~~~~~~~\simeq 1-\prod_{l\neq m}\left(1 -
\rho_T\rho_S \left\langle\sigma_{i,j}^{(l,m)}\right\rangle\right),
\label{meanedge}
\end{eqnarray}
where in the last term we take advantage of neglecting correlations  
by replacing the  average 
of the product of variables with the product of the
averages and using  Eq.~(\ref{aver}). This 
expression simply states that each possible source-target pair 
weights in the average with the product of the probability that the
end nodes are a source and a target; the discovery probability is
thus obtained by considering the edge in an average effective medium
({\em mean-field}) of sources and targets homogeneously distributed in the
network. This approach is indeed akin to
mean-field methods customarily used in the study of many particle
systems where each particle is considered in an effective average  
medium defined by the uncorrelated averages of quantities. 
The realization average of $\left\langle\sigma_{i,j}^{(l,m)}\right\rangle$ 
is very simple in the uncorrelated picture, depending only of the kind 
of $\mathcal{M}$-path. In the case of the 
ASP probing we have $\left\langle\sigma_{i,j}^{(l,m)}
\right\rangle=\sigma_{i,j}^{(l,m)}$,
in that each path contributes to the discovery of the edge. 
In
the case of the USP and the RSP, however, only one path among all the
equivalent ones is chosen, and in the average we have that each
shortest path gives a contribution
$\sigma_{i,j}^{(l,m)}/\sigma^{(l,m)}$
to
$\left\langle\sigma_{i,j}^{(l,m)}\right\rangle$,
where $\sigma^{(l,m)}$ is the number of equivalent shortest path
between vertices $l$ and $m$.

The standard situation we consider is the one in which 
$\rho_T\rho_S\ll 1$ and since 
$\left\langle\sigma_{i,j}^{(l,m)}\right\rangle\leq 1 $, we have  
\begin{equation}
\prod_{l\neq m}\left(1 -\rho_T\rho_S \left\langle\sigma_{i,j}^{(l,m)}\right\rangle\right)\simeq
\prod_{l\neq m}\exp \left(-\rho_T\rho_S
\left\langle\sigma_{i,j}^{(l,m)}\right\rangle\right),
\end{equation}
that inserted in Eq.(\ref{meanedge}) yields 
\begin{eqnarray}
\nonumber\left\langle\pi_{i,j}\right\rangle \simeq 1-
\prod_{l\neq m} \left(\exp \left(- \rho_T\rho_S
\left\langle\sigma_{i,j}^{(l,m)}\right\rangle\right)\right)\\
= 1-\exp
\left(-\rho_T\rho_S b_{ij}
\right),
\label{edgedisc}
\end{eqnarray}
where
$b_{ij}=\sum_{l\neq m}\left\langle\sigma_{i,j}^{(l,m)}\right\rangle$. In
the case of the USP and RSP probing, the quantity $b_{ij}$ is by definition
the edge betweenness centrality \cite{freeman77,brandes}, 
sometimes also refereed to as ``load''~\cite{goh01}
(In the case of ASP probing, it is a closely related quantity).   
Indeed the vertex or edge betweenness is defined as the total 
number of shortest paths among pairs of vertices in the network that 
pass through a vertex or an edge, respectively. If there are 
multiple shortest paths between a pair of vertices, the path 
contributes to the betweenness  with the corresponding relative
weight.
The betweenness gives a measure of the amount of all-to-all traffic that
goes through an edge or vertex, if the shortest path is used as the 
metric defining the optimal path between pairs of vertices,
and it can be considered as a non-local measure 
of the \textit{centrality} of an edge  or vertex in the graph.

The edge betweenness assumes values between $2$ and $N(N-1)$ and the
discovery probability of the edge will therefore depend strongly on
its betweenness. In particular, for vertices with minimum 
betweenness \mbox{$b_{ij}=2$} we have 
 \begin{equation}
\left\langle\pi_{i,j}\right\rangle \simeq 2\rho_T\rho_S ,
\end{equation}
that recovers the probability that the two end vertices of the edge
are chosen as source and target. This implies that if the densities 
of sources and targets are small but finite in the limit of very large
$N$, all the edges in the underlying graph have an appreciable
probability to be discovered. Moreover, for a large majority of edges
with high betweenness the discovery probability approaches one 
and we can reasonably expect to have a fair sampling of the network.

In most realistic samplings, however, we face a very different
situation. 
While it is reasonable to consider $\rho_T$ a small but finite value, the
number of sources is not extensive ($N_S\sim \mathcal{O}(1)$) and their 
density tends to zero as $N^{-1}$. In this case it is more convenient
to express the edge discovery probability as 
\begin{equation}
\left\langle\pi_{i,j}\right\rangle \simeq 
 1-\exp
\left(-\epsilon \overline{b_{ij}}
\right),
\label{edgesample}
\end{equation}
where $\epsilon=\rho_T N_S$ is the density 
of probes imposed to the system  and 
the rescaled betweenness $\overline{b_{ij}}=N^{-1}b_{ij}$ 
is now limited in the interval $[2N^{-1},N-1]$. In the limit of large
networks $N\to\infty$ it is clear 
that edges with low betweenness have 
$\left\langle\pi_{i,j}\right\rangle \sim \mathcal{O}(N^{-1})$, for 
any finite value of $\epsilon$. This readily tells us that in real
situations the discovery process is generally not complete, a large
part of low betweenness edges being not discovered, and that 
the network sampling is  made progressively more accurate by
increasing the density of probes $\epsilon$.  

A similar analysis can be performed for the discovery indicator function
$\pi_{i}$ of a
vertex $i$. For each source-target set $\Omega$ we have that 
\begin{eqnarray}
\nonumber&&\pi_{i}=1-\left(
1-\sum_{s=1}^{N_S} \delta_{i,i_s}- \sum_{t=1}^{N_T} \delta_{i,j_t}
\right)~~~~~~~~~~~~~~~~~\\
&&~~~~~~~~~\prod_{l\neq m\neq i}\left(
1 - \sum_{s=1}^{N_S} \delta_{l,i_s} 
\sum_{t=1}^{N_T} \delta_{m,j_t} \sigma_{i}^{(l,m)}
\right).
\end{eqnarray}
where $\sigma_{i}^{(l,m)}=1$ if the vertex $i$ belongs to the $\mathcal{M}$-path 
between nodes $l$ and $m$, and $0$ otherwise. 
This time it has been considered that each vertex is discovered
with probability one also if it is in the set of sources and targets.
The second term on the right hand side therefore expresses the probability that 
the vertex $i$ does not belong to the set of sources and targets and it is not
discovered by any $\mathcal{M}$-path between source-target pairs.
By using the same {\em mean-field} approximation as previously,  
the average vertex discovery probability reads as 
\begin{equation}
\left\langle\pi_{i}\right\rangle \simeq 1 - (1-\rho_S-\rho_T)
\prod_{l\neq m\neq i}\left(1 -\rho_T\rho_S
\left\langle\sigma_{i}^{(l,m)}\right\rangle\right). 
\end{equation}
As for the case of the edge discovery probability, the average
considers all possible source-target pairs weighted with probability
$\rho_T\rho_S$. 
Also in this case, each shortest path gives a contribution
$\sigma_{i}^{(l,m)}/\sigma^{(l,m)}$ to
$\left\langle\sigma_{i}^{(l,m)}\right\rangle$ for the USP and RSP 
models, while $\left\langle\sigma_{i}^{(l,m)}\right\rangle
=\sigma_{i}^{(l,m)}$ for the ASP model.
If $\rho_T\rho_S\ll 1$, by using the same approximations used to 
obtain Eq.(\ref{edgedisc}) we obtain 
\begin{equation}
\left\langle\pi_{i}\right\rangle\simeq 1 - (1-\rho_S-\rho_T)
\exp \left(-\rho_T\rho_S b_{i}
\right),
\end{equation}
where $b_{i}=\sum_{l\neq m\neq
i}\left\langle\sigma_{i}^{(l,m)}\right\rangle$. For the USP and RSP 
we have that $b_{i}$
is the vertex 
betweenness centrality that is limited in the interval $[0,N(N-1)]$
\cite{freeman77,brandes,goh01}.
The betweenness value $b_i=0$ holds for the leafs of the graph,
i.e. vertices with a single edge, for
which we recover
$\left\langle\pi_{i}\right\rangle\simeq\rho_S+\rho_T$.
Indeed, this kind of vertices are dangling ends discovered only if
they are either a source or target themselves.

As discussed before, the most usual setup corresponds to a density 
$\rho_S\sim\mathcal{O}(N^{-1})$ and in the large $N$ limit we can
conveniently write 
\begin{equation}
\left\langle\pi_{i}\right\rangle\simeq 1 - (1-\rho_T)
\exp \left(-\epsilon\overline{b_{i}}
\right),
\label{vertexsample}
\end{equation}
where we have neglected terms of order $\mathcal{O}(N^{-1})$ and 
the rescaled betweenness $\overline{b_{i}}=N^{-1}b_{i}$ is   
now defined in the interval $[0,N-1]$.
This expression points out that the probability of vertex discovery
is favored by the deployment of a finite density of targets that
defines its lower bound.

We can also provide a simple approximation for the effective
average degree $\left\langle k_i^*\right\rangle$ of the node $i$ 
discovered by our sampling process. Each edge departing from the vertex 
will contribute proportionally to its discovery probability, yielding   
\begin{equation}
\left\langle k_i^*\right\rangle= 
\sum_j \left( 1 - \exp \left(-\epsilon\overline{b_{ij}}\right) \right)
\simeq \epsilon\sum_j\overline{b_{ij}}. 
\end{equation}
The final expression is obtained for  edges with 
$\epsilon\overline{b_{ij}}\ll 1$. In this case,
the sum over all neighbors of the edge betweenness is
simply related to the vertex betweenness as 
$\sum_j{b_{ij}}= 2(b_{i}+ N-1)$, where the factor $2$ considers that
each vertex path traverses two edges and the term $N-1$ accounts for
all the edge paths for which the vertex is an endpoint. This finally
yields 
\begin{equation}
\left\langle k_i^*\right\rangle\simeq
2\epsilon + 2\epsilon\overline{b_{i}}.
\label{degdisc}
\end{equation}

The present analysis shows that the
measured quantities and statistical properties of the sampled graph strongly
depend on the parameters of the experimental setup and the topology of the 
underlying graph. The latter dependence is exploited by the key role
played by edge and vertex betweenness in the expressions
characterizing the graph discovery. The betweenness is a nonlocal
topological quantity whose properties change considerably depending on
the kind of graph considered. This allows an intuitive understanding 
of the fact that graphs with diverse topological properties deliver 
different answer to sampling experiments. 

\section{Numerical exploration of graphs}
In this section we use the analytical
results as a guidance in the discussion of extensive numerical
simulations of sampling experiments in a wide range of underlying
graphs derived from different models.

\subsection{Graph models definition}\label{models}
In the following, we will analyze sparse undirected graphs denoted
by $G=(V,E)$ where the topological properties of a graph are 
fully encoded in its
adjacency matrix $a_{ij}$, whose elements are $1$ if the edge $(i,j)$
exists, and $0$ otherwise.

In particular we will consider two main classes of graphs:
i) {\em Homogeneous graphs} in which, for large degree $k$, 
the degree distribution $P(k)$ decays exponentially or faster;
ii) {\em Scale-free graphs} for which $P(k)$ has a heavy tail
decaying as a power-law $P(k) \sim k^{-\gamma}$. 
Here the {\em homogeneity} refers to the existence of a meaningful
characteristic average degree that represents the typical value in the
graph. Indeed, in graphs with poissonian-like 
degree distribution a vast majority of vertices has degree close to
the average value and deviations from the average are exponentially
small in number. On the contrary, scale-free graphs are very
heterogeneous  with very large fluctuations of the degree, characterized
by a  variance of the degree distribution which diverges with the size of
the network.

Another important characteristic discriminating the topology of graphs
is the clustering coefficient 
\begin{equation}
c_i = \frac{1}{k_i (k_i -1)} \sum_{j,h} a_{ij}a_{ih}a_{jh} \ ,
\end{equation}
that measures the local cohesiveness of nodes. 
It indeed gives the fraction of connected neighbors of a given node
$i$. The average clustering coefficient $C = \frac{1}{N} \sum_i c_i $
provides an indication of the global level of cohesiveness of the
graph. The number is generally very small in random graphs that lack of 
correlations. In many real graphs the clustering coefficient appears
to be very high and opportune models have been formulated to
represent this property. 

In the following we will make use of those models that 
can be considered prototypical examples of the various classes.

\subsubsection{Erd\"os-R\'enyi model}
The classical Erd\"os-R\'enyi (ER) model \cite{er} for random
graphs $G_{N,p}$, consists of $N$ nodes, each edge being present 
in $E$ independently with probability $p$.
The expected number of edges is therefore $|E|=pN(N-1)/2$. In order
to have sparse graphs one thus needs to have $p$ of order $1/N$, since
the average degree is $p(N-1)$.
Erd\"os-R\'enyi graphs are a typical example of 
homogeneous graph, with degree distribution following
a Poisson law, and very small clustering coefficient (of order $1/N$).
Since $G_{N,p}$ can consist of more than one connected component, we 
consider only the largest of these components.

\subsubsection{Watts-Strogatz model}
The construction algorithm proposed by Watts and Strogatz for
small-world networks \cite{watts98} is the following: the 
graph is initially a one-dimensional lattice of length $N$, 
with periodic boundary
conditions (i.e. a ring), each vertex being connected to its $2m$
nearest neighbors (with $m > 1$). The vertices are then visited one
after the other; each link connecting a vertex to one of its $m$
nearest neighbors in the clockwise sense is left in place with
probability $1-p$, and with probability $p$ is reconnected to a
randomly chosen other vertex.  Long range connections are therefore
introduced. The number of edges is $|E|=Nm$, independently of $p$.
The degree distribution has a shape similar to the case
of Erd\"os-R\'enyi graphs, peaked around the average value.
The clustering coefficient, however,  is large if $p\ll 1$, making
this network a typical example of homogeneous but clustered network. 
As for the ER case, it is possible to obtain graphs consisting of
more than one connected component; in this case we use the largest
of these components.

\subsubsection{The Barab\'asi-Albert model}

Albert and Barab\'asi have proposed to combine two ingredients
to obtain a heterogeneous scale-free graph\cite{sf}:
i) {\em Growth:} Starting from an initial seed of $N_0$ vertices
connected by links, a
new vertex $n$ is added at each time step. This new site is connected
to $m$ previously existing vertices;
ii) {\em Preferential attachment rule:}  
a node $i$ is chosen by $n$ according to the probability
$P_{n\to i}=\frac{k_i}{\sum_j k_j}$;
i.e. with a probability proportional to its degree.
After $m$ different vertices have been chosen to be
connected to $n$, the growth process is iterated by
introducing a new vertex, i.e. going back to step i) until the
desired size of the network is reached.

This mechanism yields a connected graph of $|V|=N$ nodes
with $|E|=mN$ edges. 
Graphs constructed with these rules have two important characteristics:
the degree distribution of the nodes follow a power-law
$P(k) \sim k^{-\gamma}$ with $\gamma=3$, and 
the clustering coefficient is small.

\subsubsection{Clustered and scale-free graph model}

Dorogovtsev, Mendes and Samukhin have introduced
in Ref.~\cite{dm} a model of growing network with very large
clustering coefficient $C$: at each time step, a new node is
introduced and connected to {\em the two extremities of a randomly
chosen edge}, thus forming a triangle. A given node is thus chosen
with a probability proportional to its degree, which corresponds to
the preferential attachment rule. The graphs thus obtained have $N$
nodes, $2N$ edges, and  a large clustering coefficient ($\approx
0.74$) along with  a power-law distribution for the degree 
distribution of the nodes.

The main  properties of the various graphs are summarized in
table \ref{table}: Note that the clustering is indeed large
for the WS and the DMS models, and that, 
for the scale-free
models (BA and DMS), the maximum value of the connectivity ($k_{max}$)
is much larger than the average $\overline{k}$.

\begin{table}[b]
\caption{\small 
Main characteristics of the graphs used in the numerical exploration.
}
\begin{center}
\begin{tabular}{|c|c|c|c|c|c|}\hline
  & ER & ER& WS & BA & DMS \\
\hline
 $N$ & $10^4$ & $10^4$  &  $10^4$  &  $10^4$ &  $10^4$ \\
 $|E|$ &  $10^5$  &  $5. 10^5$ & $10^5$ &  $4. 10^4$ & $2. 10^4$ \\
 $\overline{k}$  & $20$ & $100$  &  $20$  & $8$ & $4$  \\
$C$ &  $0.002$  & $0.01$   & $0.52$  &  $0.006$  & $0.74$ \\
 $k_{max}$  &   $40$  &  $140$   &  $26$  &  $334$   & $346$ \\
\hline
\end{tabular}
\end{center}
\label{table}
\end{table}

\subsection{Sampling homogeneous graphs}

\begin{figure}[t]
\begin{center}
\includegraphics[width=8.0cm]{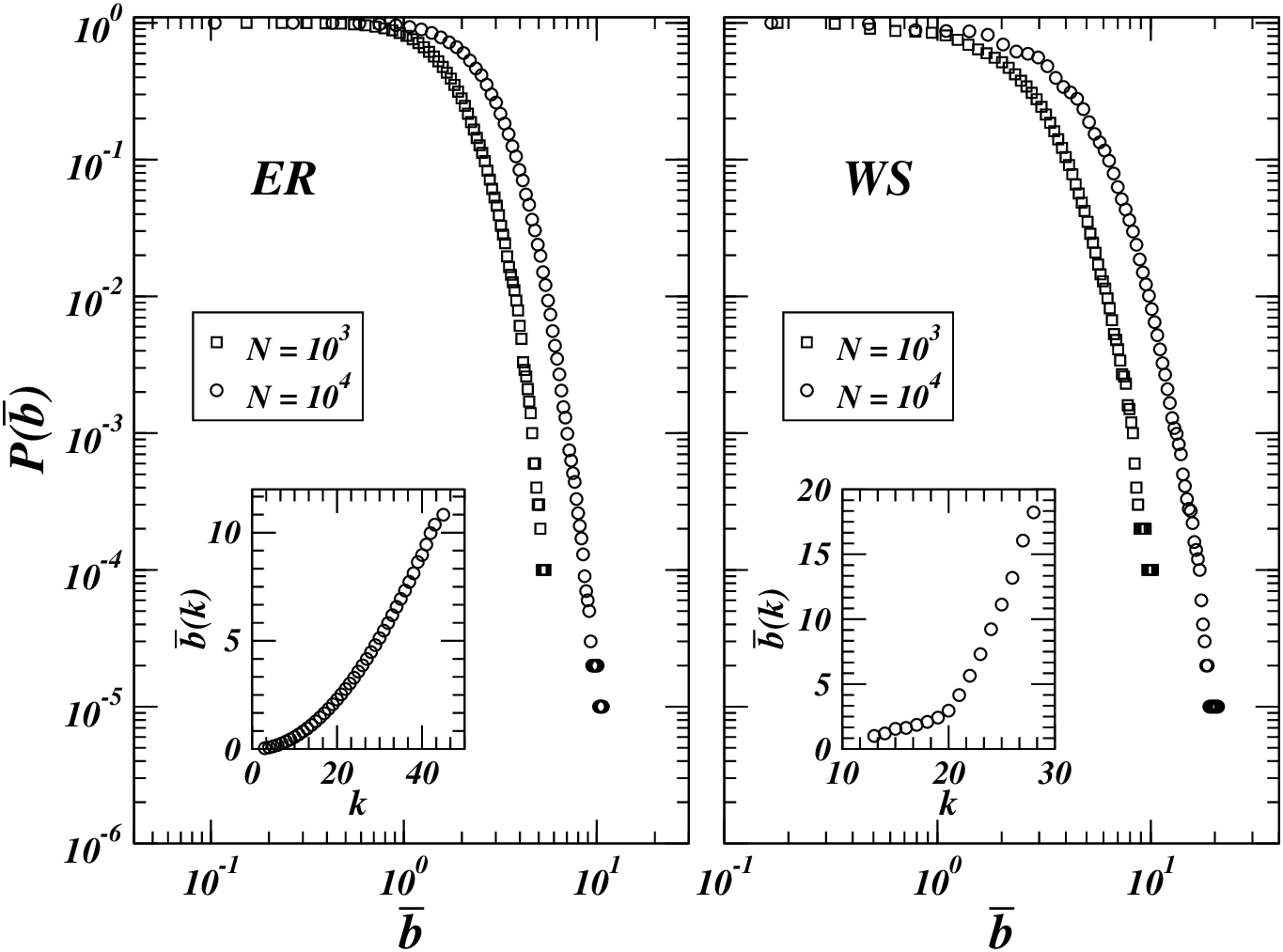}
\end{center}
\caption{Cumulative distribution of the average node betweenness $\overline{b}$
in the ER and WS graph models ($\overline{k}=20$). The inset 
(in lin-lin scale) shows the behavior of the 
average node betweenness as a function of the degree $k$.}
\label{bet1}
\end{figure}
\begin{figure}[t]
\begin{center}
\includegraphics[width=8.7cm]{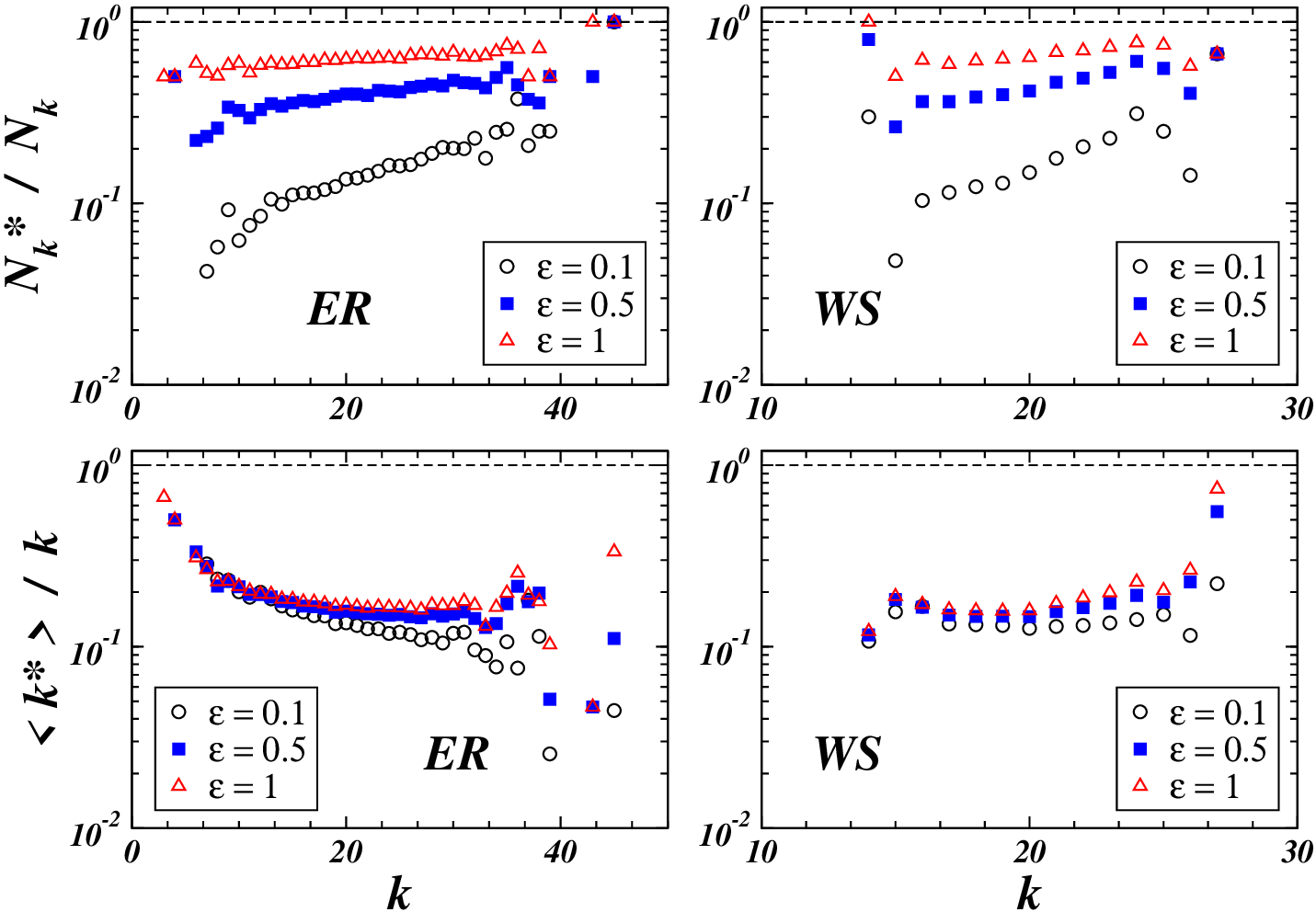}
\end{center}
\caption{Frequency $N_k^*/N_k$ of detecting a
vertex of degree $k$ (top) and proportion of discovered edges 
$\left\langle k^*\right\rangle/k$ (bottom) as a function of the degree
in the ER and WS graph models. The exploration setup considers $N_S=2$
and increasing probing level $\epsilon$ obtained by progressively
higher density of targets $\rho_T$. The $y$ axis is in log scale to
allow a finer resolution.}
\label{fig:2}
\end{figure}
Our first set of experiments consider underlying graphs with
homogeneous connectivity; namely the  Erd\"os-R\'enyi (ER) and 
the Watts-Strogatz (WS) models. 
We have used networks with $N=10^4$ nodes, $\overline{k}=20$ unless
otherwise specified; for the WS model, $p=0.1$ has been taken.
We have averaged each measurement over $10$ realizations.
Both  models have a Poissonian degree
distribution with exponential decaying tails. The distribution is
therefore peaked around the average degree $\overline{k}$
that represents the typical degree of a node. 
Since the topological properties governing the traceroute 
exploration properties is the betweenness centrality it is worth 
reviewing its general properties in the case of the models considered
here. In Fig~\ref{bet1} we report the vertex
betweenness distribution for both the ER and WS models, 
confirming their poissonian distribution with an exponentially fast 
decaying tail. The vertex and edge 
betweenness are as well homogeneous quantities in these networks and
their distributions are  peaked around the average 
values  $\overline{b}$ and 
$\overline{b_{e}}$, respectively. 
These values can 
be considered as typical values and the betweenness distribution is 
narrowly distributed around these characteristic values.
Moreover, on average, the betweenness is related to the degree of the
vertices obtaining a $\overline{b}(k)$ that increases with the degree. 
On the other hand, in homogeneous graphs the range of variation in the
degree is extremely limited, reverberating in small variations of
the betweenness values. Finally, it must be noted that the degree and 
betweenness distributions do not exhibit a pronounced scaling with 
the size of the network because of the intrinsic exponential cut-off.

Since a large majority of vertices and edges will have a betweenness
very close to the average value, we can use Eq.~(\ref{edgesample}) and 
(\ref{vertexsample}) to estimate the order of magnitude of
probes that allows a fair sampling of the graph. Indeed, both
$\langle\pi_{i,j}\rangle$ and $\langle\pi_{i}\rangle$ tend to $1$ if 
$\epsilon\gg$ max$\left[\overline{b}^{-1},
\overline{b_{e}}^{-1}\right]$. In this limit 
all edges and vertices will have probability to be discovered very
close to one. 
\begin{figure}[t]
\begin{center}
\includegraphics[width=8.0cm]{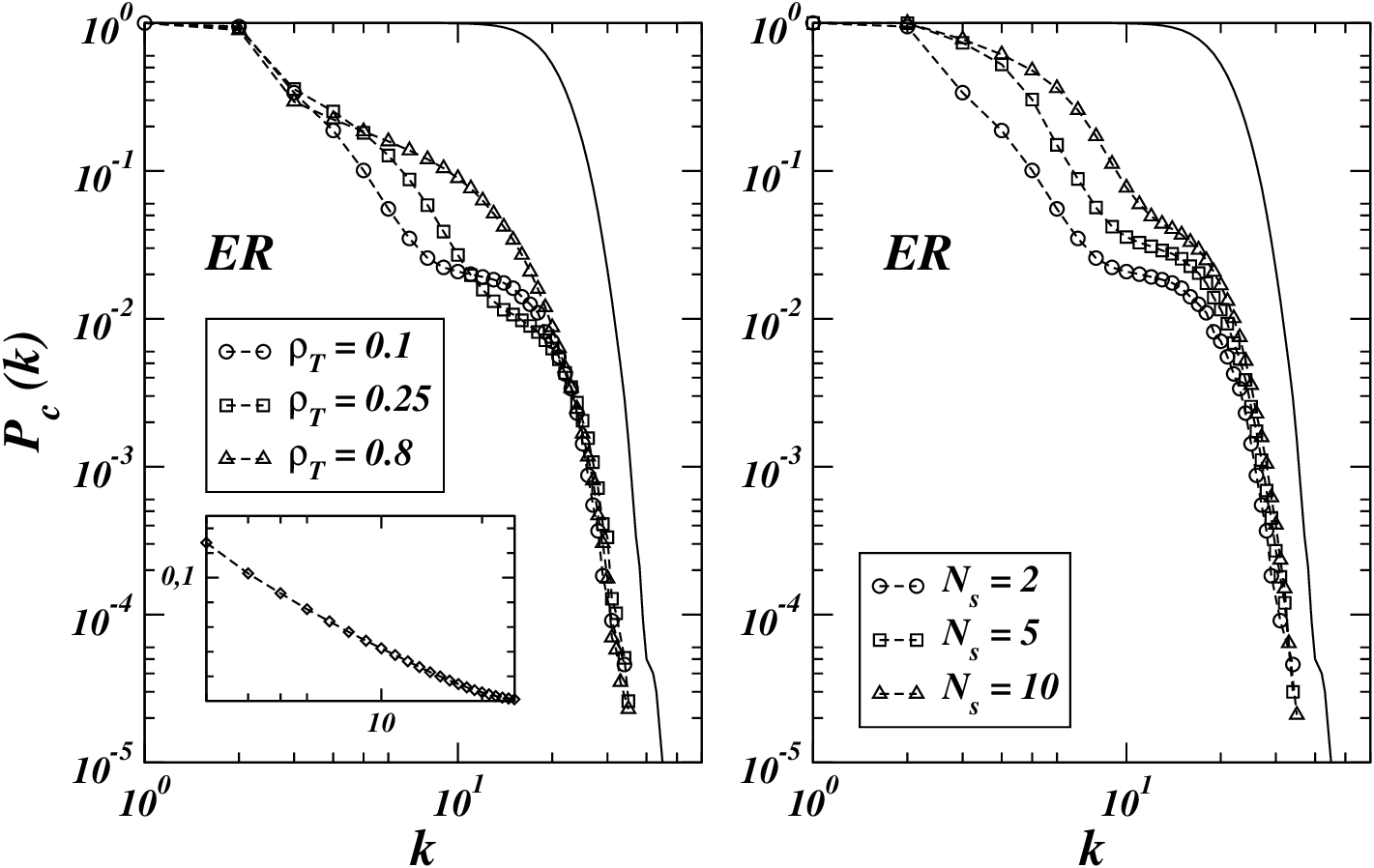}
\end{center}
\caption{Cumulative degree distribution of the sampled ER graph 
with $\overline{k}=20$ for
USP probes.  
The figure on the left shows sampled distributions obtained with $N_S=2$ 
and varying density target $\rho_T$. In the inset we report the
peculiar case $N_S=1$ that provides an apparent power-law behavior 
with exponent $-1$ at all values of $\rho_T$. The inset is in lin-log
scale to show the logarithmic behavior of the corresponding 
cumulative distribution.
The right figure shows sampled distributions obtained with
$\rho_T=0.1$ and varying number of sources $N_S$.The solid line is the 
degree distribution of the underlying graph.}
\label{fig:3}
\end{figure}

At lower value of $\epsilon$, obtained by varying $\rho_T$ and $N_S$,
the underlying graph is only partially discovered. 
We first studied the behavior of the fraction $N_k^*/N_k$  
of discovered vertices of degree $k$, where $N_k$ is the total number of
vertices of degree $k$ in the underlying graph, and the fraction of discovered 
edges $\left\langle k^*\right\rangle/k$ in vertices of degree $k$.   
In Fig.~\ref{fig:2} we report the behavior of these quantities as a
function of $k$ for both the ER and WS models. The
fraction $N_k^*/N_k$ naturally increases by augmenting the density of
targets and sources, and it is slightly increasing for larger degrees.
The latter behavior can be easily understood by noticing that vertices
with larger degree have on average a larger betweenness $b(k)$. 
By using Eq.(\ref{vertexsample}) we have that 
$N_k^*/N_k\sim 1-\exp\left(-\epsilon\overline{b(k)}\right)$, obtaining the
observed increase at large $k$. On the other hand, the range of
variation of degrees in homogeneous graphs is very narrow and only a
large level of probing may guarantee very large discovery
probabilities.  Similarly the behavior of the effective 
discovered degree can be understood by looking at
Eq.~(\ref{degdisc}) stating that $\left\langle
k^*\right\rangle/k\simeq \epsilon k^{-1}(1 + \overline{b(k)})$. Indeed
the initial decrease of $\left\langle
k^*\right\rangle/k$ is finally  compensated by the increase of
$\overline{b(k)}$. 

\begin{figure}[t]
\begin{center}
\includegraphics[width=8.0cm]{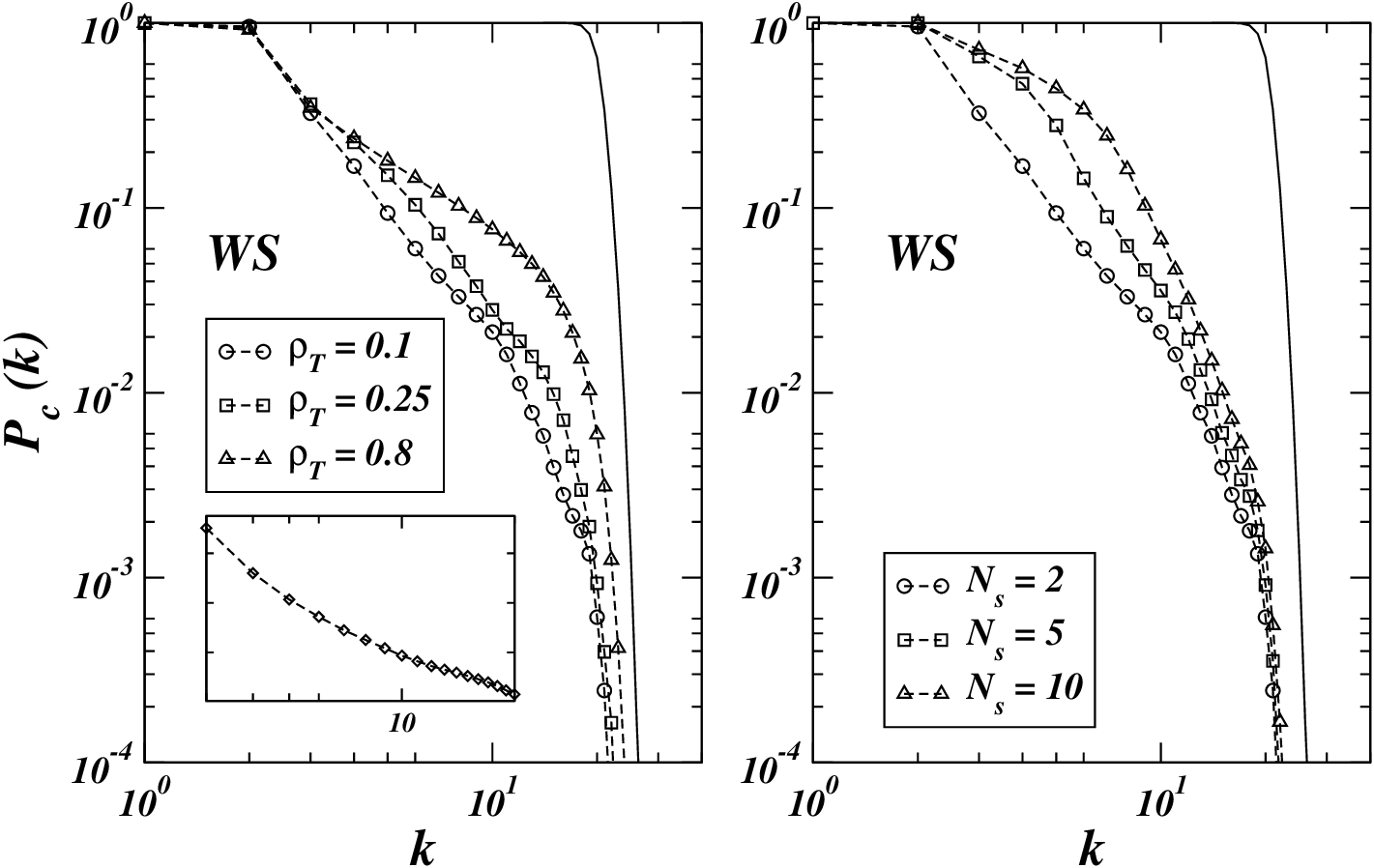}
\end{center}
\caption{Cumulative degree distribution of the sampled WS graph for
USP probes.  
The left figure shows sampled distributions obtained with $N_S=2$ 
and varying density target $\rho_T$. 
The figure on the right shows sampled distributions obtained with
$\rho_T=0.1$ and varying number of sources $N_S$.The solid line is the 
degree distribution of the underlying graph. The inset shows the logarithmic 
behavior of the cumulative distribution for $N_S=1$ and $\rho_T=1$.}
\label{fig:4}
\end{figure}

\begin{figure}[t]
\begin{center}
\includegraphics[width=8.7cm]{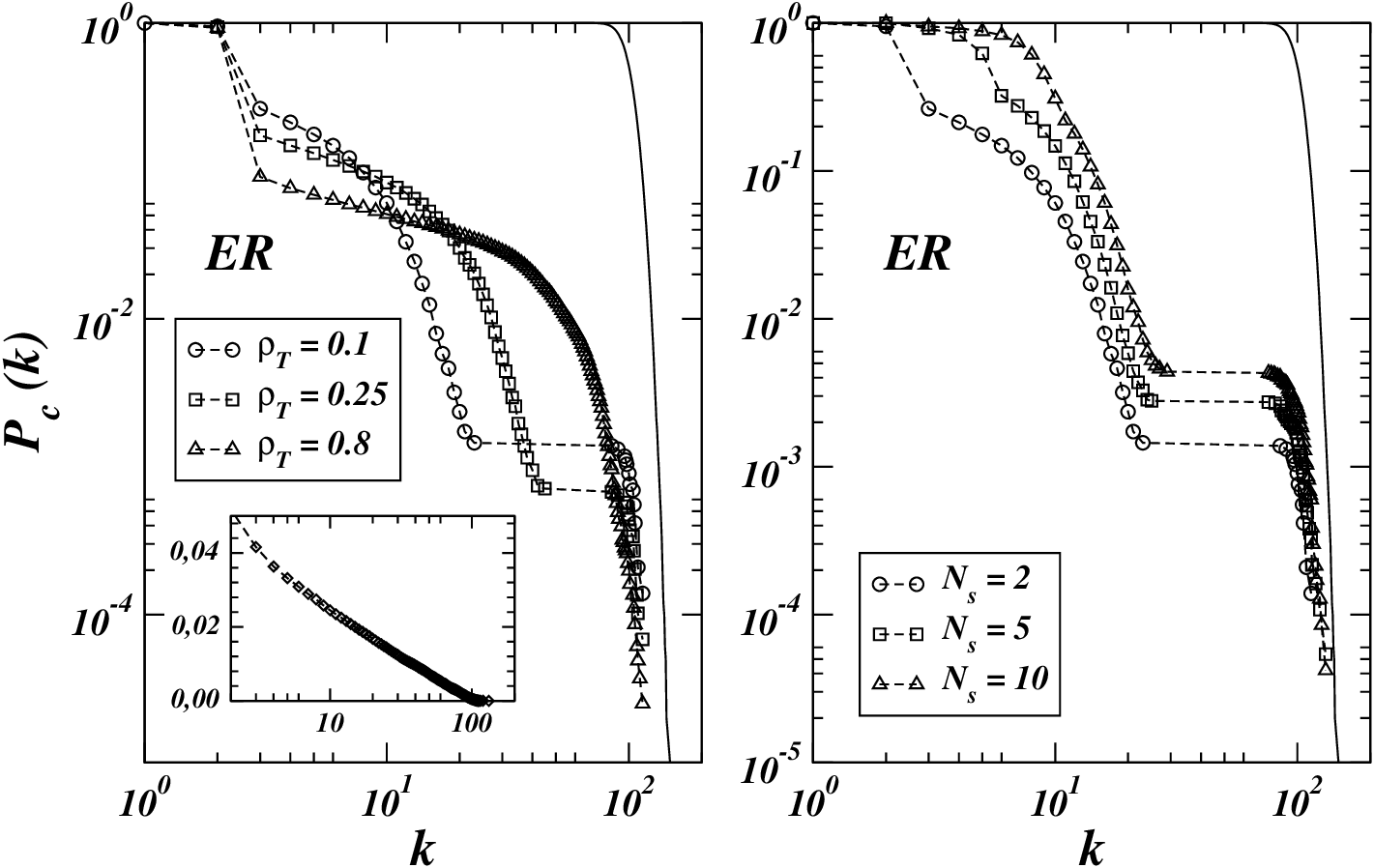}
\end{center}
\caption{Cumulative degree distribution of the sampled ER  with 
$\overline{k}=100$.  Left figure: $N_S=2$, and
various values of $\rho_T$. The inset corresponds to $N_S=1$ as in figure
\ref{fig:3}. Right figure: $\rho_T=0.1$, various values of $N_S$.  
In these cases the distribution
shows the distinctive presence of plateaus corresponding to the peaks induced 
by the sampling process.}
\label{fig:5}
\end{figure}

A very important quantity in the study of the statistical accuracy
of the sampled graph is the degree distribution. In Fig.~\ref{fig:3} we
show the cumulative degree distribution $P_c(k^*>k)$ 
of the sampled graph defined by the ER model for increasing 
density of targets and sources. 
Sampled distributions are only approximating the genuine distribution,
however, for $N_S\geq 2$ they are far from true heavy-tail distributions  
at any appreciable level of probing.
Indeed, the distribution runs generally over a small range of degrees,  
with a cut-off that sets in at the average degree
$\overline{k}$ of 
the underlying graph. In order to stretch the
distribution range, homogeneous graphs with 
very large average degree $\overline{k}$ must be
considered, however, other distinctive spurious effects appear in this case.
In particular, since the best sampling occurs around the high degree
values, the distributions develop peaks that show in the cumulative
distribution as plateaus (see Fig.\ref{fig:5}). 
The very same behavior is obtained in the 
case of the WS model (see Fig.~\ref{fig:4}). 
Finally, in the case of RSP and ASP  model, we observe that the 
obtained distributions are closer to the real one since they 
allow a larger number of discoveries.

Only in the  peculiar case of $N_S=1$ an apparent scale-free behavior
with slope $-1$ is observed for all target densities $\rho_T$, as
analytically shown by Clauset and Moore \cite{clauset}. 
Also in this case, the distribution cut-off is consistently determined
by the average degree $\overline{k}$.
It is worth noting that the experimental setup with a single source is
a limit case corresponding to a highly asymmetric probing process; it is
therefore badly, if at all, captured by our statistical analysis which
assumes homogeneous deployment.

The present analysis shows that in order to obtain a sampled 
graph with apparent scale-free behavior on a degree range 
varying over $n$ orders of magnitude we would need the very 
peculiar sampling of a homogeneous underlying graph with an average 
degree $\overline{k}\simeq 10^n$; 
a rather unrealistic situation in the
Internet and many other information systems where $n\geq 2$. 

\subsection{Sampling scale-free graphs}

In this section, we extend the analysis made for homogeneous
graphs to the case of highly heterogeneous scale-free graphs. 
We consider the Barab\'asi-Albert (BA) and the 
Dorogovtsev, Mendes and Samukhin (DMS) graph models defined in section 
\ref{models}. We have used networks of size $N=10^4$ with $\overline{k}
=8$ for BA and $\overline{k}=4$ for the DMS, and
averaged each measurement over $10$ realizations.
Both models have a scale-free distribution $P(k)\sim k^{-\gamma}$ with 
$\gamma\simeq 3$. Moreover, the DMS model is highly clustered with an
average $C \simeq 0.74$. The average degree of both models is well defined,
however, the degree distribution is heavy-tailed with fluctuations
diverging logarithmically with the graph size. This implies that
$\overline{k}$ is not a typical value in the 
network and there is an appreciable
probability of finding vertices with very high degree.  
\begin{figure}[t]
\begin{center}
\vspace{.7cm}
\includegraphics[width=8.7cm]{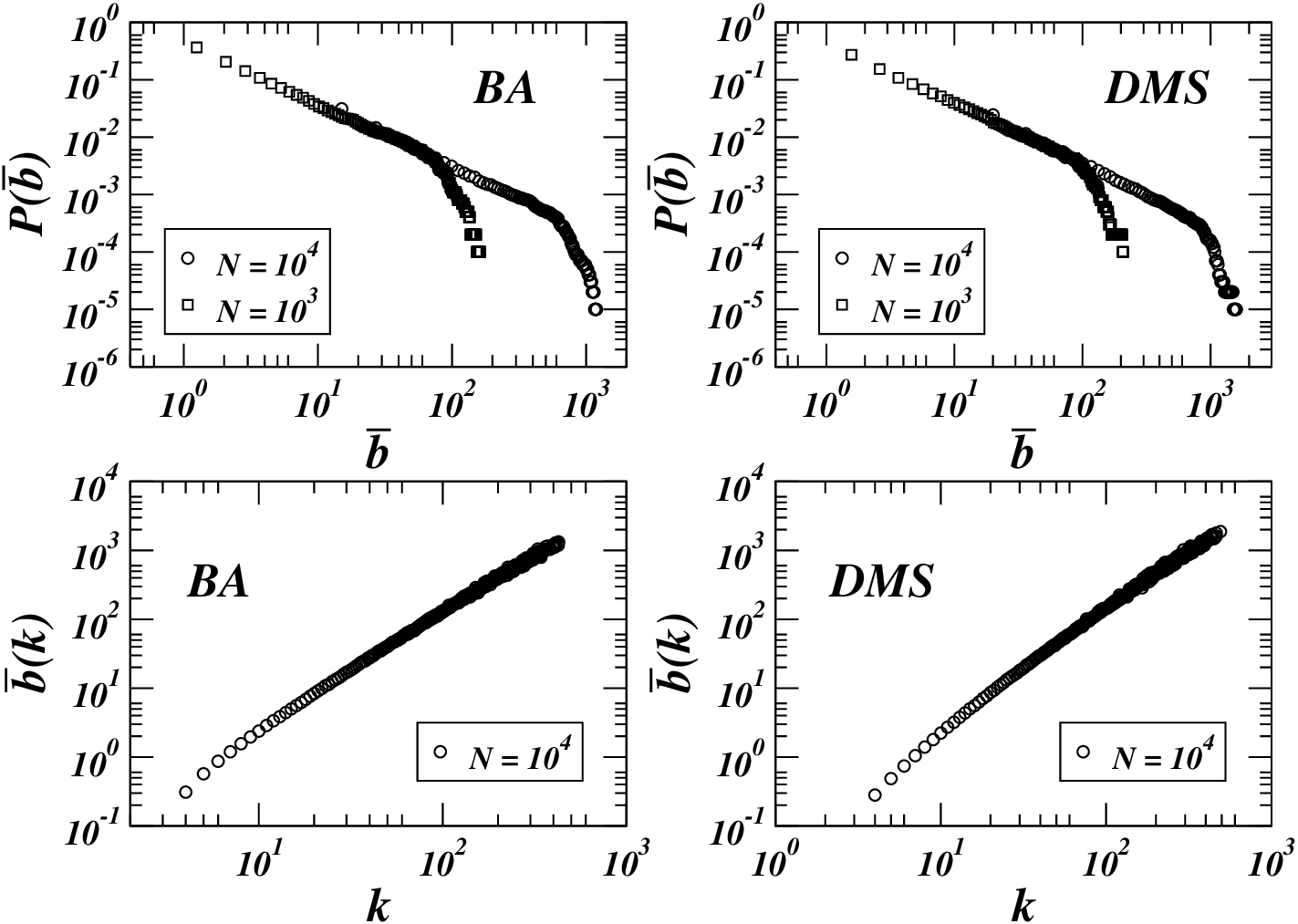}
\end{center}
\caption{Cumulative distribution of the average node betweenness $\overline{b}$ 
(top) and its behavior as a function of the degree $k$ (bottom) in the BA and
DMS graph models. The plot is in log-log scale.}
\label{bet2}
\end{figure}
\begin{figure}[t]
\begin{center}
\includegraphics[width=8.7cm]{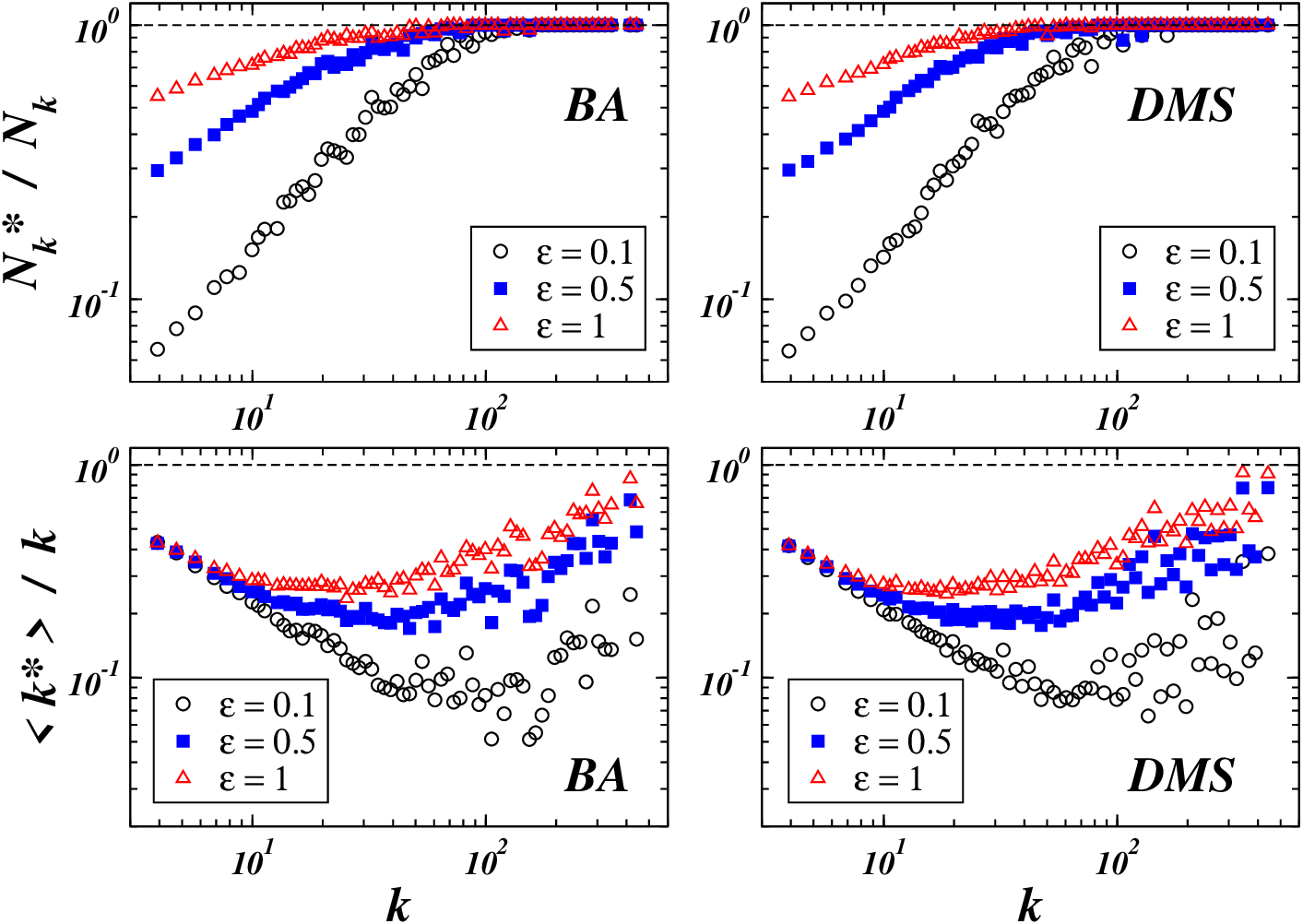}
\end{center}
\caption{Frequency $N_k^*/N_k$  of detecting a
vertex of degree $k$ (top) and proportion of discovered edges 
$\left\langle k^*\right\rangle/k$ (bottom) as a function of the degree
in the BA and DMS graph models. The exploration setup considers $N_S=2$
and increasing probing level $\epsilon$ obtained by progressively
higher density of targets $\rho_T$. The plot is in log-log scale to
allow a finer resolution and account for the wide variation of degree
in scale-free graphs.}
\label{fig:6}
\end{figure}
Analogously, the betweenness distribution is heavy-tailed,
allowing for an appreciable fraction of vertices and edges with very high 
betweenness\cite{bart}.  
In particular it is possible to show (see Fig.~\ref{bet2}) 
that in scale-free graphs the site betweenness is related to the vertices 
degree as $\overline{b(k)}\sim k ^\beta$, where $\beta$ is an exponent 
depending on the model \cite{bart}. Since in heavy-tailed 
degree distributions the allowed degree is varying over several orders 
of magnitude, the same will
occur for the betweenness values, as shown in Fig.~\ref{bet2}. 
In addition, as customary for
scale-free graphs, the betweenness distribution extends on a range of
values that increases with the size of the network: i.e. in principle
it does extend up to infinity in an infinite network.

In such a situation, even in the case
of small $\epsilon$, vertices whose  betweenness is large enough
($\overline{b(k)}\epsilon\gg 1$) 
have $\left\langle \pi_i\right\rangle\simeq 1$. Therefore all vertices with
degree $k\gg \epsilon^{-1/\beta}$ will be detected with probability
one. This is clearly visible in Fig.~\ref{fig:6} where the discovery 
probability $N_k^*/N_k$ of vertices with degree $k$ saturates to one for
large degree values. Consistently, the degree value at which the curve
saturates decreases with increasing $\epsilon$. 
A similar effect is appearing in the measurements concerning 
$\left\langle k^*\right\rangle/k$. After an initial decay 
(see Fig.~\ref{fig:6}) the effective discovered degree is increasing 
with the degree of the vertices. This qualitative feature is captured
by Eq.~(\ref{degdisc}) that gives  $\left\langle
k^*\right\rangle/k\simeq \epsilon k^{-1}(1 + \overline{b(k)})$.
After an initial decay the term $k^{-1}\overline{b(k)}\sim k^{\beta -1}$
takes over and the effective discovered degree approaches the real
degree $k$.

\begin{figure}[t]
\begin{center}
\includegraphics[width=7.5cm]{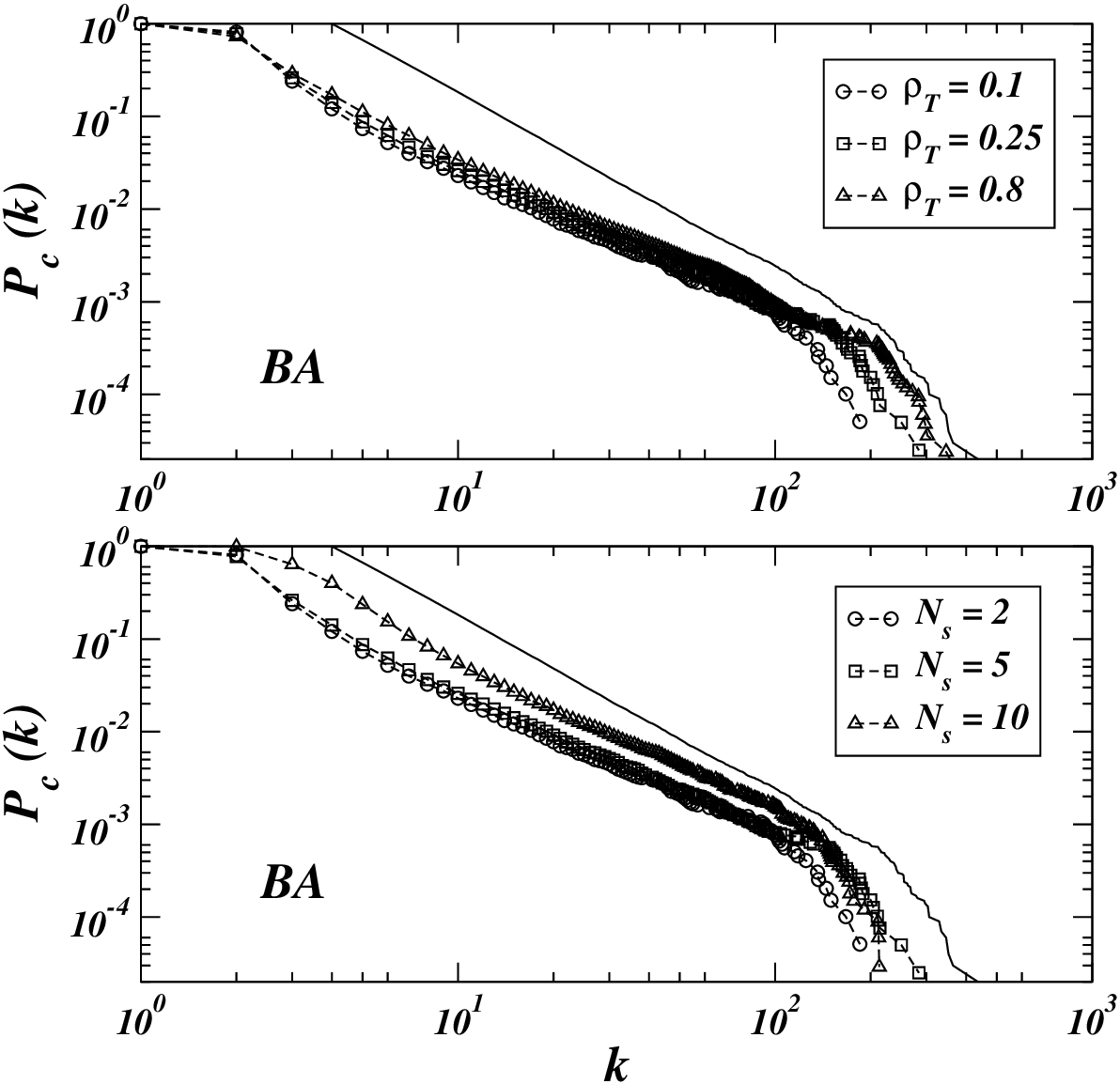}
\end{center}
\caption{Cumulative degree distribution of the sampled BA graph for
USP probes.  
The top figure shows sampled distributions obtained with $N_S=2$ 
and varying density target $\rho_T$. 
The figure on the bottom shows sampled distributions obtained with
$\rho_T=0.1$ and varying number of sources $N_S$. The solid line is the 
degree distribution of the underlying graph.}
\label{fig:7}
\end{figure}
\begin{figure}[t]
\begin{center}
\includegraphics[width=8.0cm]{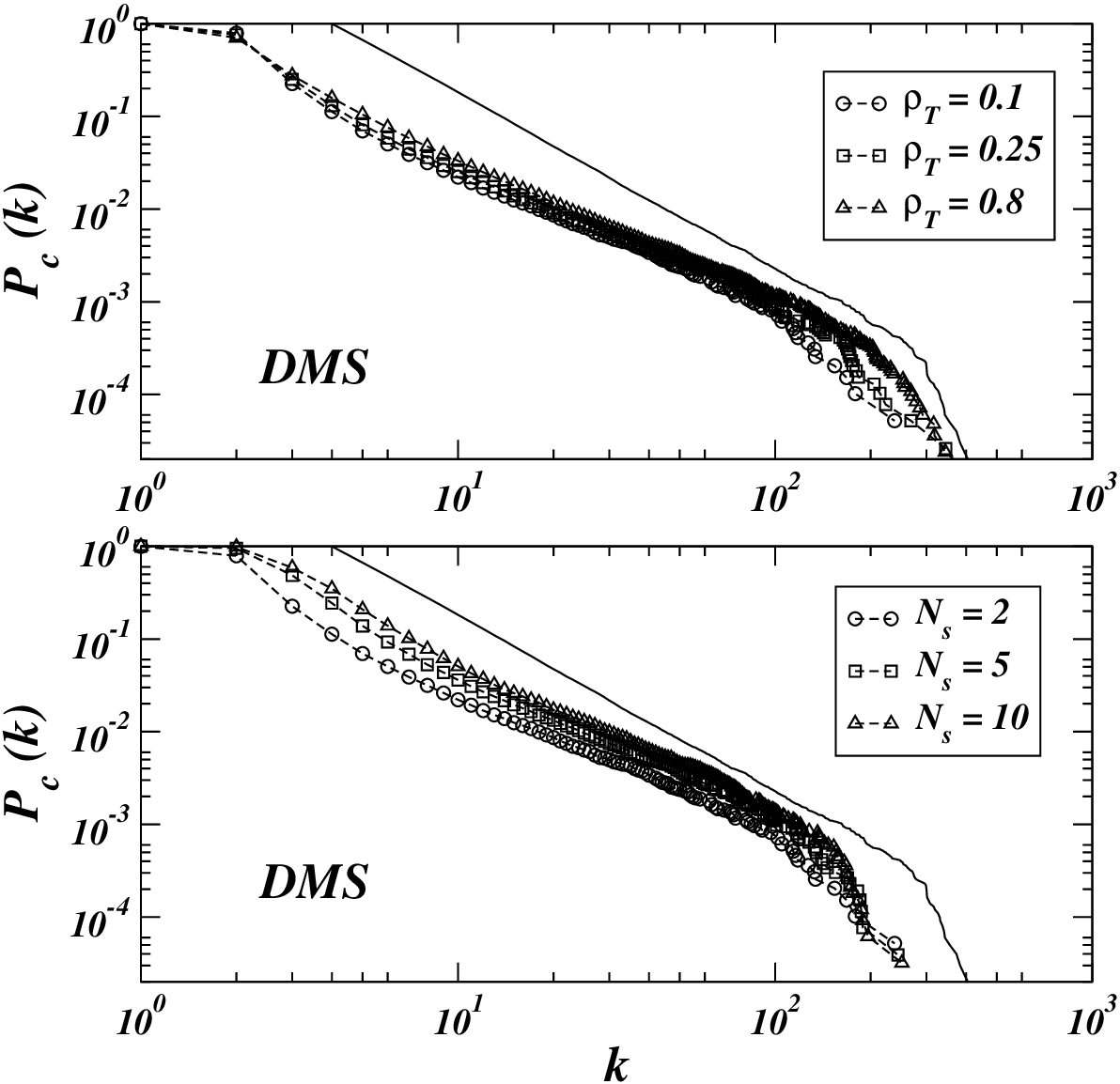}
\end{center}
\caption{Cumulative degree distribution of the sampled DMS graph for
USP probes.  
The top figure shows sampled distributions obtained with $N_S=2$ 
and varying density target $\rho_T$. 
The figure on the bottom shows sampled distributions obtained with
$\rho_T=0.1$ and varying number of sources $N_S$. The solid line is the 
degree distribution of the underlying graph.}
\label{fig:8}
\end{figure}

It is evident from the previous discussions, that in scale-free
graphs,  vertices with high degree are efficiently sampled 
with an effective measured degree that is rather close to the real
one. This means that the degree distribution tail is fairly well sampled
while  deviations should be expected at lower degree values.
This is indeed what we observe in numerical experiments on BA and DMS
graphs (see Figs.~\ref{fig:7} and \ref{fig:8}). 
Despite both underlying graphs have a small average degree,
the observed degree distribution spans more than
two orders of magnitude. The distribution tail is fairly reproduced
even at rather small values of $\epsilon$. The data shows clearly that
the low degree regime is instead under-sampled providing an apparent
change in the exponent of the degree distribution. This effect has
been noticed also by Petermann and De Los Rios in the
case of single source experiments \cite{delos} . 

The present analysis points out that graphs with heavy-tailed degree
distribution allow a better qualitative representation of their
statistical features in sampling experiments. Indeed, the most
important properties of these graphs are related to the heavy-tail
part of the statistical distributions that are indeed well
discriminated by the \texttt{traceroute}-like exploration.

\section{Optimization of mapping strategies}

In the previous sections we have shown that it is possible to have a
general qualitative understanding of the efficiency of network
exploration and the induced biases on the statistical properties. 
The quantitative analysis of the sampling strategies, however, is a
much harder task that calls for a detailed study of the discovered 
proportion of  the underlying graph and the precise deployment of
sources and targets. In this perspective, very important quantities
are the fraction $N^*/N$ and $E^*/E$ of vertices and edges discovered
in the sampled graph, respectively. Unfortunately, the mean-field
approximation breaks down when we aim at a quantitative representation
of the results. The neglected correlations are in fact very important
for the precise estimate of the various quantities of interest. 
For this reason we performed an extensive set of numerical
explorations aimed at a fine determination of the level of sampling
achieved for different experimental setups.

\begin{figure}[t]
\begin{center}
\includegraphics[width=8.0cm]{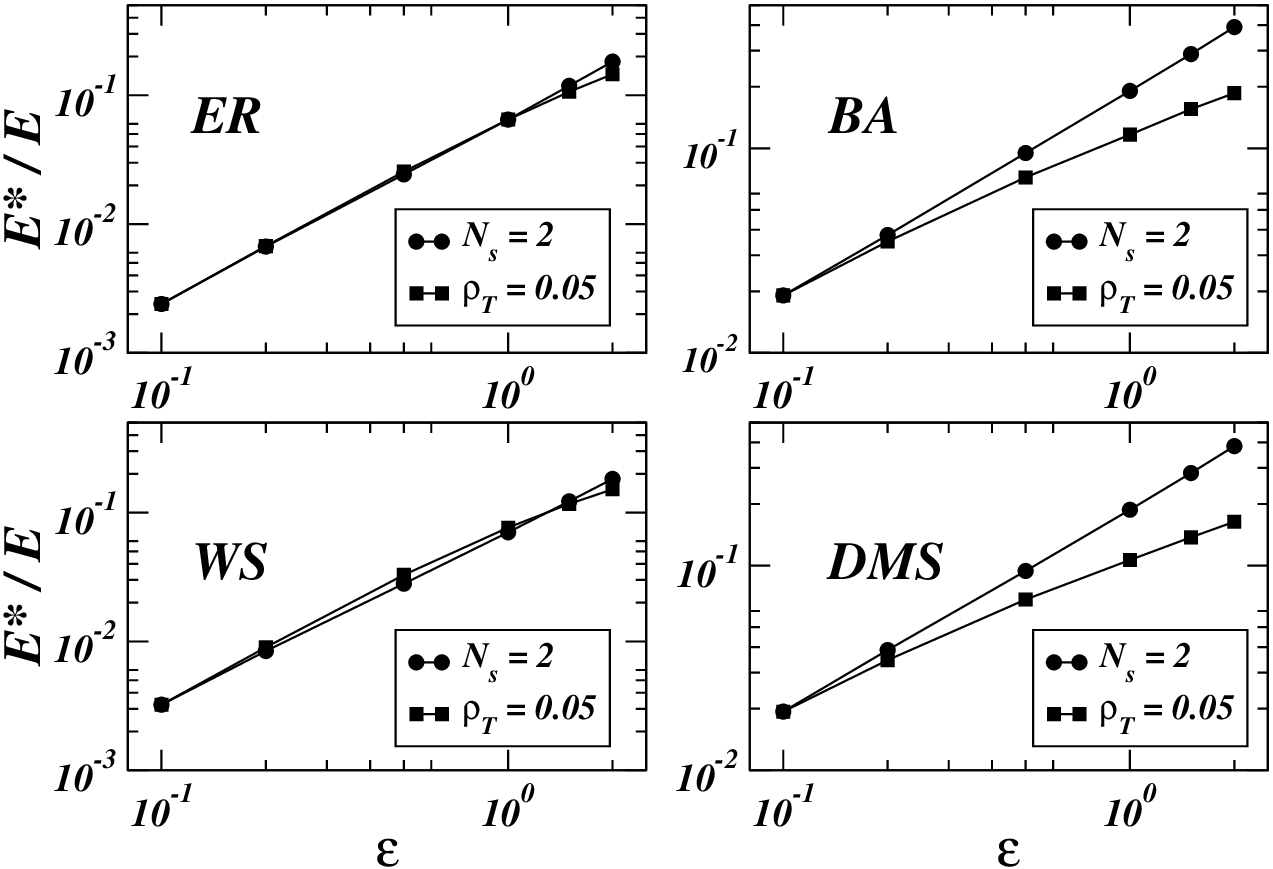}
\end{center}
\caption{Behavior of the fraction of discovered edges in explorations 
with increasing $\epsilon$. For each underlying graph studied we 
report two curves corresponding to larger $\epsilon$ achieved by increasing
the target density $\rho_T$ at $N_S=2$ or the number of sources $N_S$
at $\rho_T=0.05$.}
\label{fig:9}
\end{figure}

In Fig.~\ref{fig:9} we report the proportion of discovered edges in
the numerical exploration of the graph models defined previously for
increasing level of probing $\epsilon$. The level of probing is
increased either by raising the number of  sources at fixed target density
or by raising the target density at fixed number of sources.
As expected, both strategies  are progressively more efficient with
increasing levels of probing.  
In scale-free graphs, it is also possible to see that when the 
number of sources is $N_S\sim
\mathcal{O}(1)$ the increase of the number of targets achieves better
sampling than increasing the deployed sources. On the other hand, it
is easy to perceive that the shortest path route mapping is a
symmetric process if we exchange sources with targets. This is
confirmed by numerical experiments in which we use a very large
number of sources and a number of targets  $\rho_T\sim
\mathcal{O}(1/N)$, where the trends are opposite: the increase of 
the number of sources achieves better sampling than increasing the 
deployed targets.

\begin{figure}[t]
\vspace{-.9cm}
\begin{center}
\includegraphics[width=7.8cm,angle=270]{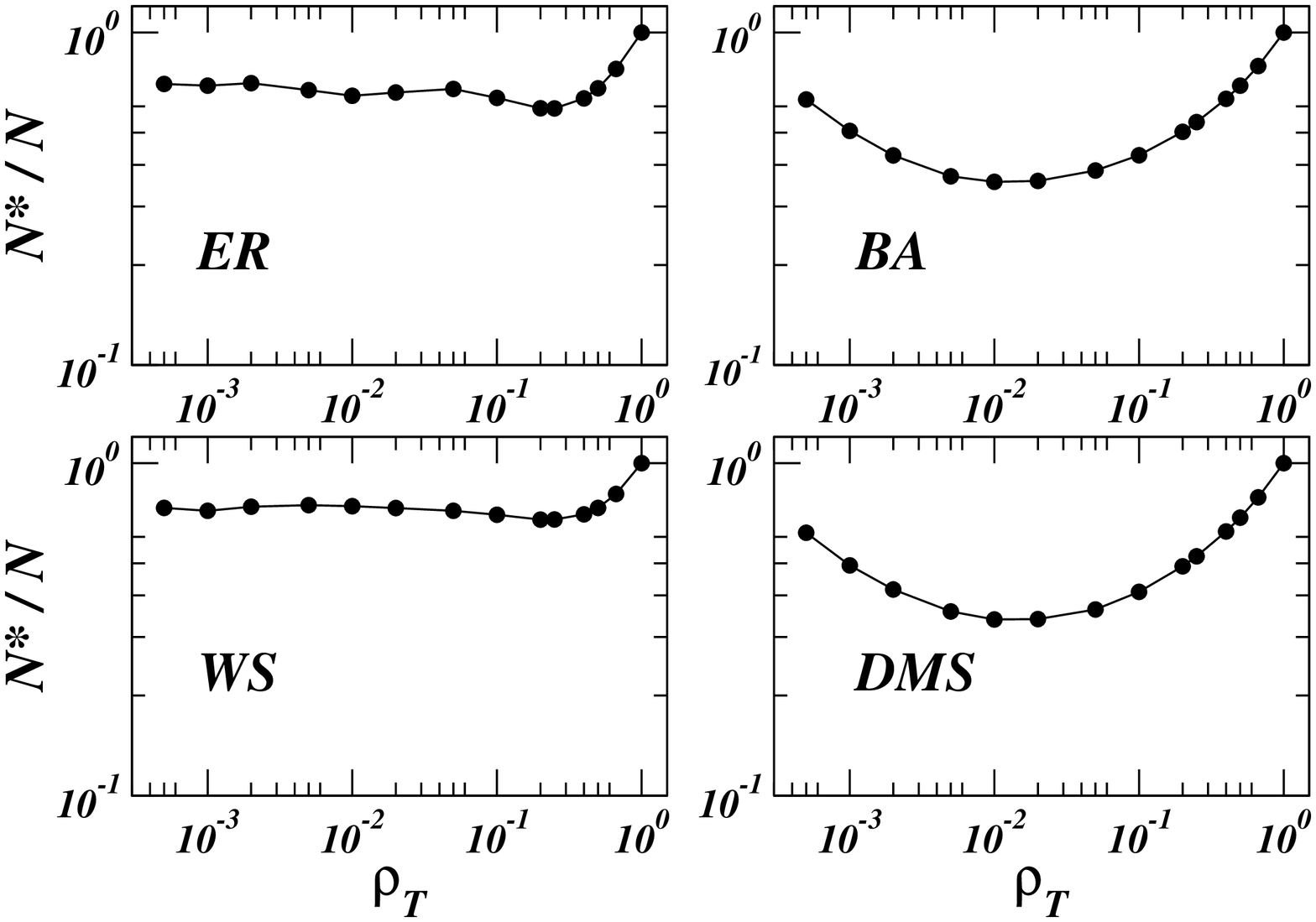}
\end{center}
\vspace{-0.6cm}
\caption{Behavior as a function of $\rho_T$ 
of the fraction of discovered nodes in explorations 
with fixed $\epsilon$ (here $\epsilon=2$).}
\label{fig:10b}
\end{figure}

This finding hints toward a behavior that is determined by the
number of sources and targets, $N_S$ and $N_T$. Any quantity is 
thus a function of $N_S$ and $N_T$,
or equivalently of $N_S$ and $\rho_T$. This point is clearly
illustrated in Fig.~\ref{fig:10} and \ref{fig:10b}, 
where we report the behavior of
$E^*/E$ and $N^*/N$ at fixed $\epsilon$ and varying $N_S$ and $\rho_T$. 
The curves exhibit a non-trivial behavior and since we will work at
fixed $\epsilon=\rho_T N_S$, any measured quantity can then be written as
$f(\rho_T,\epsilon/\rho_T)=g_\epsilon(\rho_T)$.
Very interestingly, the curves show a structure allowing for local
minima  and maxima in the discovered portion of the underlying graph.
\begin{figure}[t]
\vspace{.7cm}
\begin{center}
\includegraphics[width=8.0cm]{fig10}
\end{center}
\caption{Behavior as a function of $\rho_T$ 
of the fraction of discovered edges in explorations 
with fixed $\epsilon$ (here $\epsilon=2$). 
Since $\epsilon=\rho_T N_S$, the increase of 
 $\rho_T$ corresponds to a lowering of the number of sources $N_S$.}
\label{fig:10}
\end{figure}

This feature can be explained by a simple symmetry argument. 
The model for \texttt{traceroute} is symmetric by the exchange of sources
and targets, which are the endpoints of shortest paths: an exploration with
$(N_T,N_S)=(N_1,N_2)$ is equivalent to one with  $(N_T,N_S)=(N_2,N_1)$.
In other words, at fixed $\epsilon=N_1 N_2/N$, a density of targets
$\rho_T=N_1/N$ is equivalent to a density
$\rho'_T= N_2/N$. Since $N_2=\epsilon/\rho_T$ we obtain that at
constant $\epsilon$, experiments with $\rho_T$ 
and $\rho'_T=\epsilon/(N\rho_T)$ are equivalent obtaining by symmetry
that any measured quantity obeys the equality 
\begin{equation}
g_\epsilon(\rho_T)= g_\epsilon \left( \frac{\epsilon}{N\rho_T} \right).
\end{equation}
This equation implies a symmetry point signaling  the presence of a
maximum or a minimum at $\rho_T =\epsilon/(N\rho_T)$. We therefore 
expect the occurrence of a symmetry in the graphs of 
Fig.\ref{fig:10} at $\rho_T\simeq \sqrt{\epsilon/N}$. Indeed, 
the symmetry point is clearly
visible and in quantitative good agreement with the previous
estimate in the case of scale-free graphs. On the contrary,
homogeneous underlying topology have a smooth behavior that makes
difficult the clear identification of the symmetry point. 
It must be also noticed that USP probes create a certain level of
correlations in the exploration that tends to  hide the complete
symmetry of the curves.

\begin{figure}[t]
\begin{center}
\includegraphics[width=8.0cm]{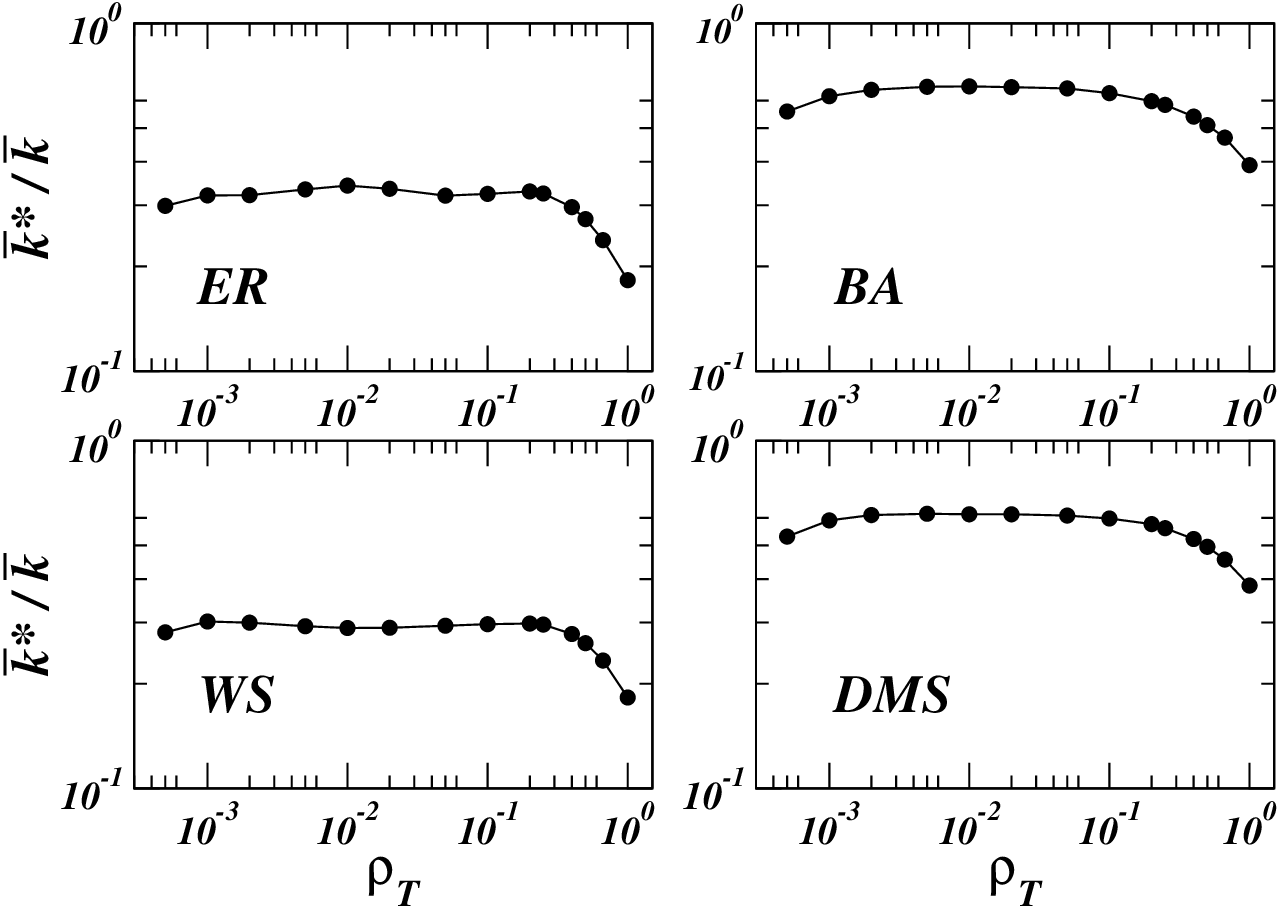}
\end{center}
\caption{Behavior as a function of $\rho_T$ 
of the fraction of the normalized average degree 
$\overline{k}^*/\overline{k}$ for 
a fixed probing level $\epsilon$ (here $\epsilon=2$).}
\label{fig:11}
\end{figure}
\begin{figure}[t]
\begin{center}
\includegraphics[width=8.0cm]{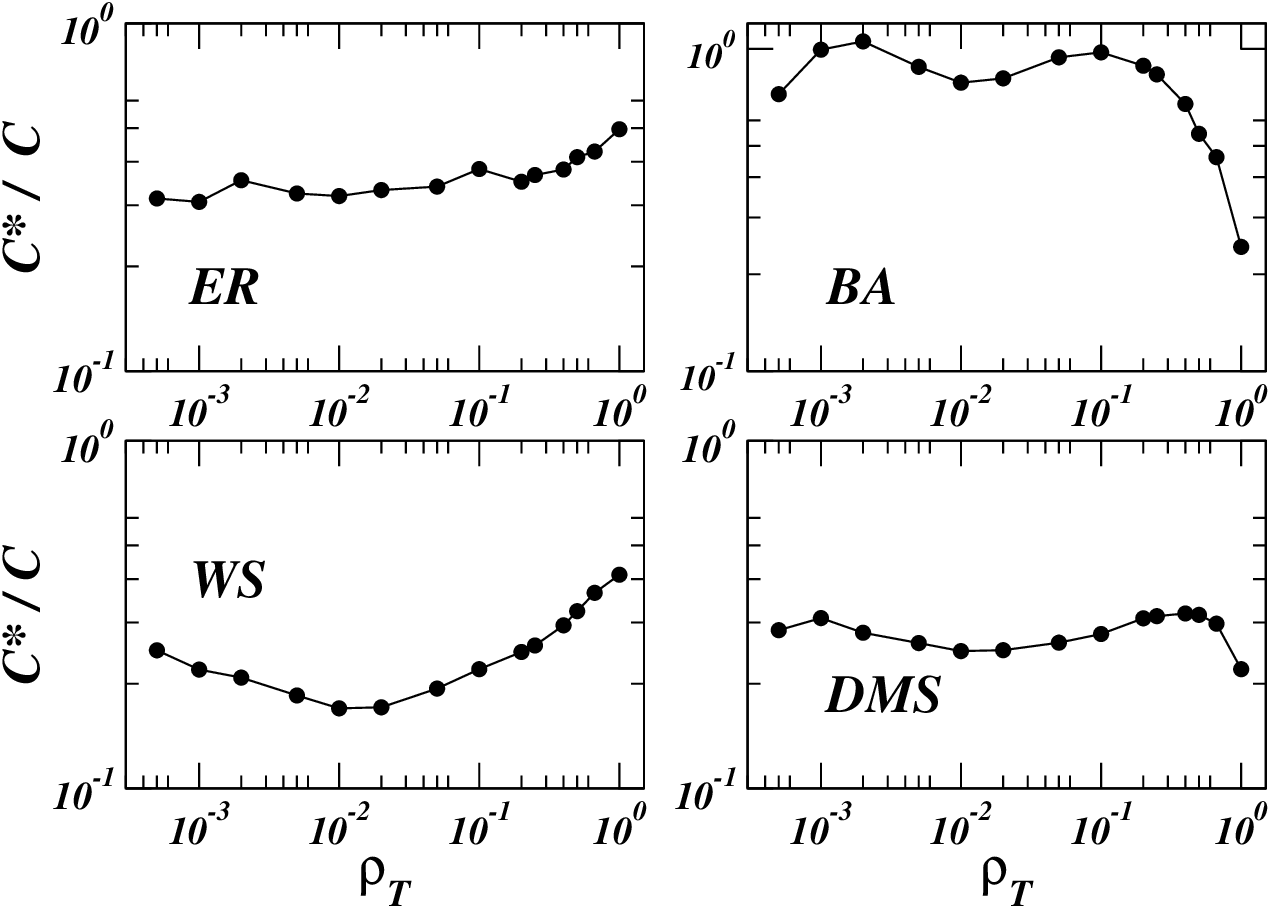}
\end{center}
\caption{Behavior as a function of $\rho_T$ 
of the fraction of the normalized average clustering coefficient 
$C^*/C$ for 
a fixed probing level $\epsilon$ (here $\epsilon=2$).}
\label{fig:12}
\end{figure}

The previous results imply that at fixed levels of probing $\epsilon$ 
different proportions of sources and targets may achieve different
levels of sampling. This hints to the search for optimal strategies in
the relative deployment of sources and targets. 
The picture, however, is more complicated if we look at other
quantities in the sampled graph. In Fig.\ref{fig:11} we show the
behavior at fixed $\epsilon$ of the average 
degree $\overline{k}^*$
measured in sampled graphs normalized with the actual average degree 
$\overline{k}$ of the underlying graph as a function of
$\rho_t$. The plot shows also in this case a
symmetric structure with a maximum at the symmetry point. By comparing  
Fig.\ref{fig:11} with Fig.\ref{fig:10} we notice that the symmetry
point is of a different nature for different quantities. 
Hence, where we have a minimum in the fraction of discovered edges, we
have the best estimate of the average degree. This implies that at the
symmetry point the exploration discovers less edges than in other setups,
however, achieving a more efficient sampling of the effective degree
for the discovered vertices. 
%
%
%
A similar problem is obtained by studying
the behavior of the ratio $C^*/C$ between the
clustering coefficient of the sampled and the underlying graphs.
We studied this quantity for the WS and DMS models that have a high
clustering level. Also in these cases, as shown in Fig.\ref{fig:12}, the 
best level of sampling is achieved at particular values of $\epsilon$
and $N_S$ that are conflicting with the best sampling of other
quantities. 
%
%

The evidence purported in this section hints to a possible optimization
of the sampling strategy. The optimal solution, however, appears as a
trade-off strategy between the different level of efficiency achieved 
in competing ranges of the experimental setup. In this respect, a 
detailed and quantitative investigation of the various quantities of 
interest in
different experimental setups is needed in order to pinpoint the most
efficient deployment of source-target pairs depending on the
underlying  graph topology.

\section{Conclusions and outlook}

The rationalization of the exploration biases at the statistical 
level provides a general interpretative framework for the results 
obtained from the numerical experiments on graph models. The sampled
graph clearly distinguishes the two situations defined by homogeneous and
heavy-tailed topologies, respectively. This is due to the exploration process
that statistically focuses on high betweenness nodes, thus providing a very
accurate sampling of the distribution tail. In graphs with heavy-tails,
such as scale-free networks, the main topological features are
therefore easily discriminated since the relevant statistical
information is 
encapsulated in the degree distribution tail which is fairly well
captured. Quite surprisingly, the sampling
of homogeneous graphs appears more cumbersome than those of
heavy-tailed graphs. Dramatic effects such as the existence 
of apparent power-laws, however, are found only in very peculiar cases. In
general, exploration strategies provide sampled distributions with 
enough signatures to distinguish
at the statistical level between graphs with different topologies.

This evidence might be relevant in the discussion of real data from
Internet mapping projects. Indeed, data available so far
indicate the presence of heavy-tailed degree distribution both at the
router and AS level. In the light of the present discussion, it is
very unlikely that this feature is just an artifact of the mapping
strategies. The upper degree cut-off at the router and AS level 
runs up to $10^2$ and $10^3$, respectively. A homogeneous graph should
have an average degree comparable to the measured cut-off and this is
hardly conceivable in a realistic perspective (for instance, it would 
require that nine routers over ten would have more than 100 links to 
other routers). In addition, the major part of mapping projects are
multi-source, a feature that we have shown to readily wash out the  
presence of spurious power-law behavior. On the contrary, power-law
tails are easily sampled with particular accuracy for the large degree
part, generally at all probing levels. This makes very plausible, and
a natural consequence, that the heavy-tail behavior observed in 
real mapping  experiments is a genuine feature of the Internet.

On the other hand, it is important to stress that while at the
qualitative level the sampled graphs allow a good discrimination of
the statistical properties, at the quantitative level they might
exhibit considerable deviations from the true values
such as average degree, distribution exponent and clustering
properties. In this respect, it is of major importance to define 
strategies that optimize the estimate of the various parameters and
quantities of the underlying graph. In this paper we have shown that
the proportion of sources and targets may impact on the accuracy of
the measurements even if the number of total probes imposed to the
system is the same. For instance, the
deployment of a highly distributed  infrastructure of sources
probing a limited number of targets may result as efficient as a few
very powerful sources probing a large fraction of the addressable space.
The optimization of large network sampling is therefore an open
problem that calls for further work aimed at a more quantitative
assessment of the mapping strategies both on the analytic and
numerical side. 

\section*{Acknowledgments}

We are grateful to M. Crovella, P. De Los Rios, T. Erlebach, T. Friedman,
M. Latapy and T. Petermann for  very useful discusssion and
comments.  This work has been partially supported by 
the European Commission Fet-Open
project COSIN IST-2001-33555 and contract 001907 (DELIS).

\end{document}